\documentclass[10pt,a4paper,notitlepage]{article}
\pdfoutput=1

\usepackage{hyperref}
\usepackage[utf8]{inputenc}
\usepackage[english]{babel}
\usepackage[T1]{fontenc}
\usepackage{empheq}
\usepackage{amsmath}
\usepackage{amsfonts}
\usepackage{caption}
\usepackage{color} 
\newcommand{\Tr}{\mathrm{Tr}}

\newtheorem{definition}{Definition}
\newtheorem{proposition}{Proposition}

\newcommand{\inv}{\mathrm{inv}}
\newcommand{\sym}{\mathrm{Sym}}

\usepackage{multicol}
\usepackage{amssymb}
\usepackage{graphicx}
\usepackage[left=1.8cm,right=1.8cm,top=2cm,bottom=2.4cm]{geometry}

\begin{document}

\begin{center}
\textbf{{\large Functional Renormalization Group Approach for Tensorial Group Field Theory: \\ 
a Rank-$6$ Model with Closure Constraint}}
\end{center}
\begin{center}
\vspace{20pt}
Dario Benedetti\footnote{dario.benedetti@th.u-psud.fr}
 and 
Vincent Lahoche\footnote{vincent.lahoche@th.u-psud.fr}\\
\vspace{5pt}
{\it Laboratoire de Physique Th\'eorique, CNRS-UMR 8627, Universit\'e Paris-Sud 11, 91405 Orsay Cedex, France}
\end{center}
\vspace{10pt}

\begin{abstract}
We develop the functional renormalization group formalism for a tensorial group field theory with closure constraint, in the case of a just renormalizable model over $U(1)^{\otimes 6}$, with quartic interactions. The method allows us to obtain a closed but non-autonomous system of differential equations which describe the renormalization group flow of the couplings beyond perturbation theory. The explicit dependence of the beta functions on the running scale is due to the existence of an external scale in the model, the radius of $S^1\simeq U(1)$.
We study the occurrence of fixed points and their critical properties in two different approximate regimes, corresponding  to the deep UV and deep IR. Besides confirming the asymptotic freedom of the model, we find also a non-trivial fixed point, with one relevant direction.
Our results are qualitatively similar to those found previously for a rank-3 model without closure constraint, and it is thus tempting to speculate that the presence of a Wilson-Fisher-like fixed point is a general feature of asymptotically free tensorial group field theories.
\end{abstract}

\setlength{\columnseprule}{1pt}
\setlength{\columnsep}{30pt}
\vspace{10pt}
\bigskip
\tableofcontents
\pagebreak
\section{Introduction}

Among the many results obtained with the renormalization group, two of the most famous are certainly the asymptotic freedom of QCD and the non-trivial IR behavior of scalar field theories below the critical dimension, due to the presence of a Wilson-Fisher fixed point.
However, the two phenomena are hard to find in the same system: the non-trivial IR behavior of QCD is due to confinement, not to an IR fixed point, and the marginal couplings of scalar field theories do not go to zero in the UV, running instead into Landau poles. One example of coexistence is provided by gauge theories with $N_f$ massless fermions, with $N_f$ close to the critical number above which asymptotic freedom is lost: in such case one finds the so-called Banks-Zaks fixed point \cite{Banks:1981nn}. However, such a fixed point is not a generic feature of gauge theories, it exists only for a restricted range of parameters. 
More recently, a new class of field theories has emerged which appears to enjoy asymptotic freedom in quite some generality, and for which we will show here another case of its coexistence with a Wilson-Fisher fixed point. Such theories are known as Tensorial Group Field Theories (TGFTs).
	
TGFTs are a particular case of Group Field Theories (GFTs) \cite{GFTreviews}, a class of field theories defined on a group manifold, whose non-local interactions give a complex cellular structure to the Feynman's graphs of the perturbation theory. Such a structure is particularly interesting for quantum gravity, because it provides a tool for summing over manifolds.
In their simplest form, and historically their original one, they appear as generalizations of matrix models \cite{Di Francesco:1993nw} to higher dimensions, in the form of {\it tensor models} \cite{tensor}.
The most fruitful form of these models are the colored tensor models, for which an expansion in $1/N$, similar to the one for matrices, has been constructed by Gurau and his collaborators \cite{expansion1,expansion2,expansion3,Bonzom:2011zz,uncoloring}. Such results boosted a fast expansion in this direction (e.g. \cite{var-tens} references therein). 
Proper GFT models can be seen as enrichments of tensor models with group-theoretic data. They originated in the context of topological field theory \cite{boulatov}, and they were further developed because of their links with loop quantum gravity \cite{Rovelli:1993,GFT-LQG} and spin-foam models \cite{Reisenberger:2001}.
Lastly, TGFTs can be seen either as an enrichment of the colored tensor models by the addition of group variables, or alternatively as GFTs whose interactions show the same structure and the same unitary invariance as the interactions of the colored tensor models. Such particular models will be studied in this paper, by means of a specific example. 

From the quantum gravity point of view, one of the main challenges of GFTs is to understand how the quantum degrees of freedom organize to form a geometric structure which can be identified with a semi-classical space-time (a challenge that GFT share with loop quantum gravity \cite{Dittrich:2014ala}). 
In this quest, the physics of phase transitions seems to be a promising route. In the most widely admitted scenario \cite{Oriti:2007qd}, and according to recent works \cite{Gielen:2013kla,Gielen:2013naa}, these phase transitions should correspond to the condensation of quanta corresponding to the fields involved in the GFTs. This phenomenon, very similar to the Bose-Einstein condensation, implies the spontaneous acquisition of a nonzero value by the mean field (something similar has been explored recently also in tensor models \cite{Delepouve:2015nia,Benedetti:2015ara}). In this approach, the help of the renormalization machinery, which perfectly matches with the GFTs formalism, is invaluable \cite{Rivasseau-track}. The renormalization of GFTs \cite{GFTrenorm} and TGFTs \cite{TGFTrenorm-Joseph,TGFTrenorm-Carrozza,TGFTrenorm-others,Lahoche:2015ola} has been studied for several years,  and the TGFTs (with non-trivial propagator) are the only models found so far which allow to build rigorously a renormalization program, a success mainly due to the loosening of the ultra-locality of usual GFTs, and to the tensorial nature of the interactions. This leads to a new notion of locality, called ``traciality'', which enables one to clearly define the contraction notion of a high graph \cite{TGFTrenorm-Carrozza}. For now, several approaches have been followed, especially perturbative ones, and some characteristics of these models start to emerge. One of them is their asymptotic freedom, which has been discovered by Ben Geloun and his collaborators by perturbative calculations \cite{TGFTrenorm-Joseph}, and which seems to be a generic characteristic of TGFTs \cite{Rivasseau-AF}.
Another aspect is the presence of non-trivial fixed points, which started being explored only recently by means of the Functional Renormalization Group (FRG) \cite{BBGO,Geloun:2015qfa} and the $\epsilon$-expansion \cite{Carrozza:2014rya}.

The FRG is a widely used formulation of the Wilsonian RG, with a vast range of applications \cite{Berges:2000ew,Bagnuls:2000ae,Delamotte-review,Rosten-review,Blaizot:2012fe,Gies:2006wv,Reuter:2012id}. For many purposes its most convenient formulation is in terms of the effective average action, which satisfies what is often called the Wetterich equation \cite{Wetterich:1992yh,Morris:1993qb}. The main advantage of the FRG is that it allows to explore different approximation methods that do not necessarily rely on having a small coupling constant, and thus are particularly interesting in the search of non-trivial fixed points. 
It is precisely for such reason that we will employ here the FRG machinery, following the path opened by its application to matrix models \cite{EichhornKoslowski} and TGFTs  \cite{BBGO,Geloun:2015qfa}.

In this paper, we will further develop the FRG approach to TGFTs, by studying the case of a just renormalizable Abelian TGFT \cite{Lahoche:2015ola}. 
After briefly reviewing TGFTs in section \ref{Sec:TGFTs}, and the FRG construction at the beginning of section \ref{Sec:FRG}, we will obtain, by the end of section \ref{Sec:FRG}, a closed system of differential equations which describe the renormalization group flow of the model. 
In sections \ref{Sec:large-s} and \ref{Sec:small-s}, we will study the occurrence of fixed points and phase transitions in two different approximate regimes, corresponding  to the deep UV and deep IR, or to the limit in which the radius of the group manifold goes to  infinite and zero, respectively. 
We will summarize and discuss our findings in section \ref{Sec:concl}, leaving the connection to the perturbative calculations for the appendices.

\section{Tensorial Group Field Theories}
\label{Sec:TGFTs}
\subsection{Generalities}

A tensorial group field theory on  $d$ copies of the compact group $G$ is built along the lines of uncolored tensor models of rank  $d$ \cite{uncoloring}.
The theory is defined by a choice of Gaussian measure $d\mu_{C}(\bar{\psi},\psi)$ and of interaction $S_{int}[\bar{\psi},\psi]$ in the generating function
\begin{equation}\label{partfunc}
\mathcal{Z}[\bar{J},J]:=\int d\mu_{C}(\bar{\psi},\psi)e^{-S_{int}[\bar{\psi},\psi]+\langle\bar{J},\psi\rangle+\langle\bar{\psi},J\rangle},
\end{equation}
where $C$ refers to the covariance of the free theory. The fields $\psi$ and $\bar{\psi}$, as well as the sources $J$ and $\bar{J}$ are functions on $G^d$ that take values in $\mathbb{C}$,
 $$\psi: G^d \to \mathbb{C}\qquad J:G^d \to \mathbb{C},$$ 
and the notation $\langle\cdot,\cdot\rangle$ means

\begin{equation}
\langle\bar{J},\psi\rangle := \int d^d[g] \bar{J}(g_1,...,g_d)\psi(g_1,...,g_d),
\end{equation}
where $d[g]$ stands for the Haar measure on the group $G$ being considered. Obviously, if $G$ has dimension $D$, the TGFT is a field theory in dimension $D\times d$. We denote by $\mathcal{T}$ and $\mathcal{\bar{T}}$ the set of fields $\psi$ and $\bar{\psi}$, respectively. The true peculiarity of TGFT models lies in the non-local form of the interactions contained in $S_{int}$, which are built as follows.
The convolution of the fields follows the outline of index contractions from the interactions of the uncolored tensor models \cite{uncoloring}: 
group elements are contracted (i.e. identified and integrated over the group) pairwise, such that an element appearing as the $i$-th argument of a $\psi$ is contracted with an element appearing as the $i$-th argument of a $\bar{\psi}$. 
A pattern of contracted fields is sometimes referred to as a {\it bubble}. 
What makes the bubbles non-local is both the restriction to pairwise identification and the fact that a given field can contract different arguments to different fields.
While the interactions are non-local on the group, the pairwise contraction allows the definition of a new locality principle, known as \textit{traciality}, essential for the renormalization. In the lattice gauge field theory language, traciality means essentially that, for all the leading order graphs, flat holonomies imply necessarily trivial connection \cite{TGFTrenorm-Carrozza}. In the case that $G$ is the abelian group $U(1)$, this amounts to imposing invariance under a formal unitary symmetry, the limit for $N$ going to infinity of the uncolored tensor models' $U(N)^{\otimes d}$ invariance. The interactions are formed by all the possible index contractions compatible with the $U(N)^{\otimes d}$ invariance, i.e. they are formed by the pairwise contraction described above, with no insertion of external functions or operators on the group. 
We refer to these interactions as {\it invariant traces}:
\begin{equation}
S_{int}[\bar{\psi},\psi]:=\sum_{b}\lambda_b\Tr_{b}[\bar{\psi},\psi],
\end{equation}
where $b$ labels each contraction possible, involving an arbitrary number of fields, and is associated to the coupling $\lambda_b$. Those contractions are graphically represented by bipartite colored graphs, whose vertices, black or white, are associated to the fields $\psi$ and $\bar{\psi}$, respectively, and whose lines correspond to the contracted field indices. For example, the graph of figure \ref{fig1} below,
\begin{center}
\includegraphics[scale=1]{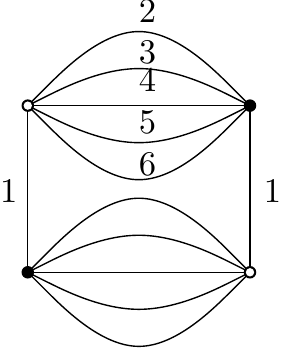} 
\captionof{figure}{Example of invariant bubble in dimension 6}
\label{fig1}
\end{center}
corresponds to the following rank-6 interaction:
\begin{align}
\int d^6[g]d^6[g']\psi(g_1,g_2,g_3,g_4,g_5,g_6)\bar{\psi}(g'_1,g_2,g_3,g_4,,g_5,g_6)\psi(g'_1,g'_2,g'_3,g'_4,g'_5,g'_6)\bar{\psi}(g_1,g'_2,g'_3,g'_4,g'_5,g'_6).
\end{align}
Other examples of bubbles are given on figure \ref{fig2}, for various ranks. 
There is obviously only one bubble constructible out of two fields, represented in the last graph of figure \ref{fig2}. This is the only local interaction in the usual sense, it corresponds to a mass term, and it is often included directly in the covariance. All the other interactions arise from the multiplication of fields at different points in $G^{\otimes d}$.

\begin{center}
\includegraphics[scale=1]{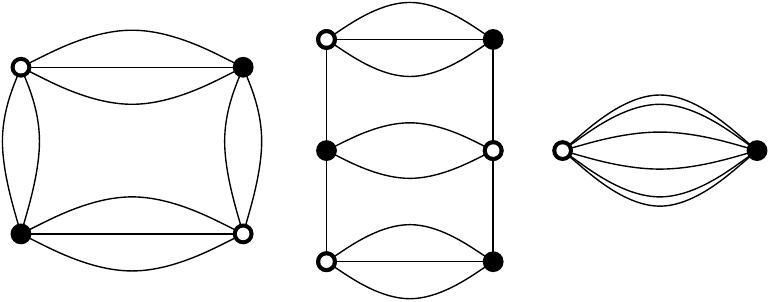} 
\captionof{figure}{Examples of interaction bubbles in dimension 5, 4 and 6}
\label{fig2}
\end{center}

In the TGFT models, a specific symmetry is often imposed at the level of the field variables, known as the closure constraint,  which reads
\begin{equation}
\psi(g_1,...,g_d)=\psi(g_1h,...,g_dh) \qquad \forall h \in G.
\end{equation}

This gauge symmetry is part of the physical enrichments added to purely tensorial models in the GFT context and it can be interpreted as geometrical data. Its origin can be found in the spin-foam foundations and LQG requirements of group field theories \cite{GFTreviews,TGFTrenorm-Carrozza}, and recent works have shown that it implies some nice properties in a purely field theoretical context. In particular, it has been pointed out that it might favor asymptotic freedom, see \cite{Rivasseau-AF}. \\

\noindent
The choice of the covariance $C$ must respect this invariance and must enforce it at the level of the Feynman graphs. The usual choice is the following, given here in Schwinger (or parametric) representation,
\begin{align} \label{propagator}
\int &d\mu_{C}(\bar{\psi},\psi)\bar{\psi}(\{g'_i\})\psi(\{g_i\}):=\int_G d[h]\int_{0}^{+\infty} d\alpha\, e^{-\alpha m^2}\prod_{i=1}^{d}K_{\alpha}(g_ih(g'_i)^{-1}),
\end{align}
where $m > 0$ is the mass parameter of the theory and $K_{\alpha}$ is the heat kernel on G, verifying
\begin{equation}
\dfrac{\partial}{\partial \alpha} K_{\alpha}=\Delta_gK_{\alpha},
\end{equation}
$\Delta_g$ being the Laplacian operator on $G$.\\

As in the standard quantum field theories, the calculation of perturbations to the Gaussian model can be evaluated thanks to the Wick theorem, and any $N$-points Schwinger function, ${S_N}$, can be evaluated as a sum over Feynman graphs. These graphs, of which figure \ref{fig3} is an example, are formed by all possible Wick contractions, represented by dashed lines, between all couples $\psi \bar{\psi}$. 

\begin{center}
\includegraphics[scale=1]{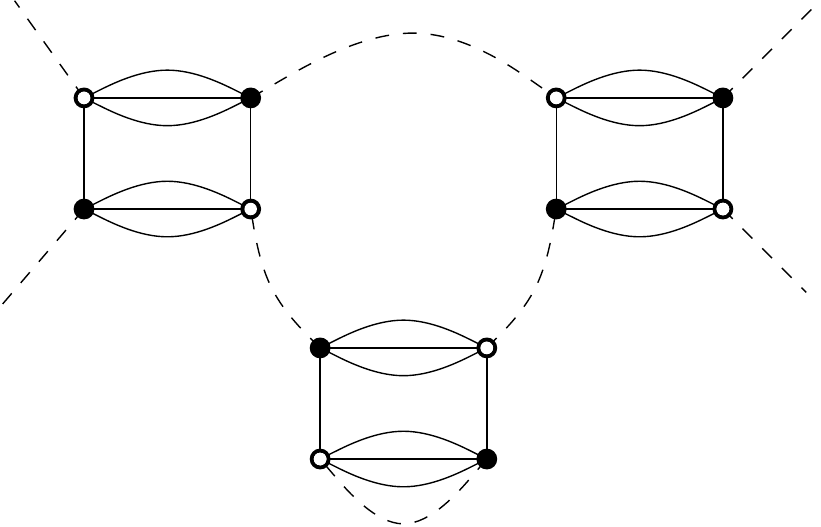} 
\captionof{figure}{A typical Feynman graph in $d=4$ with four external lines}
\label{fig3}
\end{center}

A $N$-points Schwinger function can thus be written as
\begin{equation} \label{Schwinger}
S_N=\sum_{\mathcal{G}}\dfrac{1}{s(\mathcal{G})}\left( \prod_{b\in\mathcal{G}}(-\lambda_b)\right)\mathcal{A}_{\mathcal{G}},
\end{equation}
where $s(\mathcal{G})$ is the dimension of the automorphism group of the $\mathcal{G}$ graph (the symmetry factor), $b$ label the vertices of $\mathcal{G}$ and $\mathcal{A}_{\mathcal{G}}$ is the graph amplitude. The expression of the amplitude can be obtained by using the elementary properties of the heat kernel. We find: 
\begin{align} \label{amplitude}
\mathcal{A}_{\mathcal{G}}&= {\left[ \prod_{e\in \mathcal{L}(\mathcal{G})} \int_{0}^{\infty}{d\alpha_{e} e^{-\alpha_{e} m^{2}}} \int {dh_{e}} \right]}\times{\left( \prod_{f\in \mathcal{F}(\mathcal{G})} K_{\alpha_{(f)}} {\left( \vec{\prod}_{e\in\partial{f}}h_{e}^{\epsilon_{ef}} \right)} \right)}\\\nonumber
&\qquad\qquad \qquad\qquad \qquad\times{\left( \prod_{f\in \mathcal{F}_{ext}(\mathcal{G})} K_{\alpha_{(f)}} {\left( g_{s(f)}\vec{\prod}_{e\in\partial{f}}h_{e}^{\epsilon_{ef}}g^{-1}_{t(f)} \right)} \right)}.
\end{align}
In this formula, $s(e),t(e)$ are the source and terminus of the oriented edge $e$, $\epsilon_{ef}$ designates the incidence matrix, being 0, +1 or -1 given that the line $e$ belongs or not to the boundary of $f$, the sign depending on the relative orientation, and $\mathcal{L}(\mathcal{G}), \mathcal{F}(\mathcal{G})$ and $\mathcal{F}_{ext}(\mathcal{G})$ are respectively the sets of lines, internal and external faces. Note that in the language of lattice gauge theory, the integrals over the group variables $h_e$ in the amplitude \eqref{amplitude} can be understood as a sum over discrete connections on the cellular complex defined by the Feynman graph $\mathcal{G}$. From this point of view, the product $\prod_{e\in\partial f}h_e^{\epsilon_{ef}}$ is an holonomy around the face $f$, and the $G^{\times |V(\mathcal{G})|}$ symmetry of the Feynman amplitudes (where $|V(\mathcal{G})|$ is the number of vertices in $\mathcal{G}$) coming from the invariance under the transformations
\begin{equation}
h_e\to k_{s(e)}h_ek^{-1}_{t(e)}\, , \qquad k_v\in G,
\end{equation}
is a discrete version of gauge invariance. \\

In this paper we study the case $G=U(1)$ and $d=6$, that is, a six-dimensional Abelian model,
with interactions limited to the degree 4 melonic bubbles of the same form as that of figure \ref{fig1}. All the interactions of that kind, for each possible choice of the solitary line color, are included in the model with the same coupling constant. In this model, the heat kernel has a simple expression in momentum representation 
\begin{equation}
K_{\alpha}(\theta):=\sum_{p\in\mathbb{Z}}e^{-\alpha p^2}e^{i p\theta},
\end{equation}
and the integration on the group $d[h]$ in the propagator's definition \eqref{propagator} shows a $\delta(\sum_i p_i)$, so that the expression of our model's propagator is written, in momentum representation 
\begin{equation} \label{p-propagator}
C(\vec{p},\vec{p}\,{}')=\delta_{\vec{p},\vec{p}\,{}'}\dfrac{\delta(\sum_i p_{i})}{\vec{p}\,{}^2+m^2},
\end{equation}
where $\delta(x):= \delta_{0,x}$ stands for a Kronecker delta.
The interaction part can be written as:
\begin{align}
&S_{int}[\bar{\psi},\psi]=\lambda \sum_{i=1}^{6} \Tr_{b_i}[\bar{\psi},\psi]=\sum_i \int \left(\prod_{j=1}^{4} d\vec{\theta}_j\right) \mathcal{W}_{\vec{\theta}_1,\vec{\theta}_2,\vec{\theta}_3,\vec{\theta}_4}^{(i)} \psi(\vec{\theta}_1)\bar{\psi}(\vec{\theta}_2)\psi(\vec{\theta}_3)\bar{\psi}(\vec{\theta}_4),
\end{align}
where $b_i$ designates the bubble of figure \ref{fig1} whose solitary line is of color $i$ and $\mathcal{W}^{(i)}:\mathcal{T}^2\otimes \mathcal{\bar{T}}^2\to \mathbb{C}$ is a super-tensor of order four, invariant under unitary transformations.\footnote{That is, for a fixed size $N$ of a tensor, $\mathcal{W}^{(i)} \in \inv(U(N)^{\otimes d})$} As mentioned before, these interactions are chosen to be melonic, meaning that their \textit{Gurau degree} \cite{expansion3} vanishes: $\varpi = 0$. 
In momentum space we have:
\begin{equation}\label{W_p}
\mathcal{W}^{(i)}_{\vec{p}_1,\vec{p}_2,\vec{p}_3,\vec{p}_4} = \delta_{p_{1i},p_{4i}}\delta_{p_{2i},p_{3i}}
 \prod_{j\neq i} \delta_{p_{1j},p_{2j}} \delta_{p_{3j},p_{4j}}.
\end{equation}
As explained in \cite{uncoloring}, the Gurau degree characterizes the colored graphs associated to the interaction bubble, and interestingly, it plays a role similar to the genus for the matrix models in the $1/N$ expansion of the colored tensor models. Melons appear as the leading order graphs in this framework, and it can be proved that they are dual to topological spheres.
Their names comes from the leading order graphs of the colored tensorial models, called melons in \cite{Bonzom:2011zz}. Our model is uncolored, in the sense of \cite{uncoloring}, but our Feynman graphs admit a natural colored extension, and one can show that their corresponding Gurau degree vanishes for the melonic graphs.

Because of the non-local structure of the interactions, the renormalization becomes quite non trivial. However, recent works have proved that a renormalization scheme is possible, thanks to the \textit{traciality} property of the divergent graphs (see \cite{TGFTrenorm-Carrozza}), coming from the tensorial structure of the interactions. The power counting of our model was carried out using multi-scale analysis, and it has been proven \cite{TGFTrenorm-Carrozza} that the divergent degree $\omega(\mathcal{G})$ for an amplitude $\mathcal{A}_{\mathcal{G}}$ is given by:
\begin{equation}\label{divergentdegree}
\omega(\mathcal{G})=-2L(\mathcal{G})+F(\mathcal{G})-R(\mathcal{G}),
\end{equation}
where $L(\mathcal{G})$ and $F(\mathcal{G})$ designate respectively the number of lines and faces of the graph mentioned, and $R(\mathcal{G})$ the rank of the incidence matrix defined below equation \eqref{amplitude}. It can be checked that our model is just-renormalizable, with divergent graphs of degree $2$ and $0$ requiring only mass, field and coupling renormalization. More interestingly, it can be proved that all the divergent graphs are melonic, in the sense explained before. Hence, for our model, the melonic sector contains all the graphs that need to be renormalized. 

\bigskip
\subsection{Canonical dimension of a tensorial bubble}
\label{sectiondim}
In order to investigate the influence of the dimension $d$ on the occurrence of fixed points in the FRG analysis, we need to work with dimensionless quantities, as in usual quantum or statistical field theories. In the present case, there exists an external scale $L$ (the radius of $S^1\simeq U(1)$) which we have set to 1, thus implicitly making all quantities dimensionless in the usual sense.
In fact, as pointed out in \cite{BBGO}, and as usual in the presence of an external scale \cite{Benedetti:2014gja}, this is the reason why we will obtain non-autonomous RG flow equations.
However, what matters for our purposes is how such quantities scale with the cutoff, which we refer to as \textit{scaling dimension}.
This dimensional notion appears quite naturally in the perturbative calculations. In Schwinger regularization, which introduces a cutoff $1/\Lambda^2$ in the lower end of the integral over $\alpha$ appearing in \eqref{amplitude}, we get, at the dominant order in $\Lambda$ and in dimension $d$:
\begin{align}
\Sigma_{1 \, loop,\,\infty}&=\lambda\Lambda^{d-4} K_1 \label{eqdim1}\\
\Gamma_{1 \, loop,\,\infty}^{(4)}&=24\big(-\lambda+\lambda^2 \Lambda^{d-6}K_2 \big) \label{eqdim2},
\end{align}
where $K_1$ and $K_2$ are two numerical constants independent of $\lambda, \Lambda$ or $m$ (see appendix \ref{app2}).
The exponent of $\Lambda$ in the above expressions is in general a universal number, meaning that it does not depend on the chosen renormalization (for example, we find the same exponent by regularizing the theory by means of a cutoff on the momentum).
We therefore define the (canonical\footnote{At a non-trivial fixed point the scaling dimension will in general be anomalous, i.e. different from the canonical one.}) scaling dimension $[X]$ of a quantity $X$ in such a way that by redefining $X = \bar{X} \Lambda^{[X]}$ we obtain a homogeneous expression in $\Lambda$ for the renormalization of $X$.
The equation \eqref{eqdim2} then shows that the scaling dimension of $\lambda$ is $[\lambda]=6-d$, so that both summands have the same dimension. By using this result in  \eqref{eqdim1}, we find $[\Sigma_{1 \, loop,\,\infty}]=2$, implying that $[m]=1$, since $\Sigma_{1 \, loop,\,\infty}$ gives the radiative corrections of the mass term. \\

We can easily check the coherence of these definitions at all orders, noting that the exponent of $\Lambda$ appearing in \eqref{eqdim1} and ~\eqref{eqdim2} is nothing else but the divergence degree $\omega$ of the corresponding graph. The divergent degree is given by \eqref{divergentdegree}, and it can easily checked \cite{TGFTrenorm-Carrozza} that the most divergent graphs, the so-called melons, verify:
\begin{equation}
F(\mathcal{G})-R(\mathcal{G})=(d-2)(L(\mathcal{G})-V(\mathcal{G})+1)\label{graphmelon},
\end{equation}
where $V(\mathcal{G})$ is the number of vertices in the $\mathcal{G}$ graph. 
The equation \eqref{graphmelon}, together with the combinatorial relation $L(\mathcal{G})=2V(\mathcal{G})-N_{ext}(\mathcal{G})/2$, leads to 
\begin{equation}
\omega(\mathcal{G})=(d-6)V(\mathcal{G})+\Big[(d-2)+\dfrac{4-d}{2}N_{ext}(\mathcal{G})\Big].
\end{equation}
We limit ourselves to the interactions that we can find in the initial theory. A 1PI graph with four external legs has, \textit{a priori} the same dimension as the coupling constant $\lambda$. That is to say:
\begin{align*}
(d-6+[\lambda])&\big(V(\mathcal{G})-1\big)+(d-6)+\Big[(d-2)+\dfrac{4-d}{2}\times 4\Big]=0,
\end{align*}
implying $[\lambda]=6-d$. The same argument applied to an 1PI function with two external legs justifies $[m]=1$:
\begin{equation}
(d-6+[\lambda])V(\mathcal{G})+\Big[(d-2)+\dfrac{4-d}{2}\times 2\Big]=2.
\end{equation}

\section{FRG approach for TGFT}	
\label{Sec:FRG}
In this section we apply the Functional Renormalization Group formalism to the TGFTs introduced above.
First we derive  the Wetterich equation \cite{Wetterich:1992yh,Morris:1993qb} in this context, and later we use it to extract the flow equations in a simple approximation. For general reviews and applications of the FRG we refer to  \cite{Berges:2000ew,Bagnuls:2000ae,Delamotte-review,Rosten-review,Blaizot:2012fe,Gies:2006wv,Reuter:2012id}.

\subsection{Effective average action for TGFTs}
Starting from the generating functional defined by the equation \eqref{partfunc}, we define the following deformation:
\begin{equation}\label{family}
\mathcal{Z}_{s}[\bar{J},J]:=\int d\mu_{C}(\bar{\psi},\psi)e^{-S_{int}(\bar{\psi},\psi)-\Delta S_{s}[\bar{\psi},\psi]+\langle \bar{J},\psi \rangle+\langle\bar{\psi},J\rangle},
\end{equation}
where we have added to the action an IR cutoff or ``momentum-dependent mass term'' $\Delta S_{s}$, chosen ''ultralocal'' in the momentum representation, and formally defined as:
\begin{equation}
\Delta S_{s}[\bar{\psi},\psi]:= \langle\bar{\psi}, R_{s} \psi \rangle= \sum_{\vec{p} \in \mathbb{Z}^d}R_{s}(\vec{p}) T_{\vec{p}} \bar{T}_{\vec{p}},
\end{equation}
where $T_{\vec{p}}$ is the Fourier transform (or Peter-Weyl decomposition) of the field: $\psi(\vec{\theta})=\sum_{\vec{p}\in\mathbb{Z}^d} T_{\vec{p}} e^{i\vec{p}\cdot \vec{\theta}}$. 
As usual, in order to make sense of the partition function we assume that a UV regulator is also present, e.g. a sharp cutoff on the momenta $|p|\leq \Lambda$ (for the usual norm: $|p|=\sqrt{\vec{p}\cdot\vec{p}}$). In practice, one often works in the limiting case $\Lambda\to\infty$, because the Wetterich equation is well defined in that limit (although not its path integral origin).

The cutoff function $R_{s}(\vec{p})$ is a positive definite function chosen so that:\\

$\bullet$ $R_{s}(\vec{p})\geq 0$ for all $\vec{p}\in \mathbb{Z}^d$ and $s\in(-\infty,+\infty)$.\\

$\bullet$ $\lim_{s\to-\infty} R_s(\vec{p}) =  0$, implying:
\begin{equation}
\mathcal{Z}_{s=-\infty}[\bar{J},J]=\mathcal{Z}[\bar{J},J].
\end{equation}
This condition ensures that the original model is in the family \eqref{family}. Physically, it means that the original model is recovered when all the fluctuations are integrated out.  \\

$\bullet$  $\lim_{s\to\ln\Lambda} R_s(\vec{p}) =  +\infty$, ensuring that all the fluctuations are frozen when $e^s=\Lambda$.
As a consequence, the bare action will be represented by the initial condition for the flow at $s=\ln\Lambda$.\\

$\bullet$ For $-\infty<s<\ln \Lambda$, the cutoff $R_{s}$ is chosen so that 
\begin{equation}
R_{s}(|p|>e^s)\ll 1,
\end{equation}
a condition ensuring that the UV modes $|p|> e^s$ are almost unaffected by the additional cutoff term, while $R_{s}(|p|<e^s)\sim 1$, or  $R_{s}(|p|<e^s)\gg 1$, will guarantee that the IR modes $|p|< e^s$ are decoupled.\\

$\bullet$  $\frac{d}{ds} R_s (\vec{p}) \leq 0$, for all $\vec{p}\in \mathbb{Z}^d$ and $s\in(-\infty,+\infty)$, which means that high modes should not be suppressed more than low modes.\\

\noindent
As it stands, \eqref{family} defines an infinite-dimensional deformation of the original partition function. However, the role of the precise cutoff function, chosen to satisfy the above requirements, is secondary with respect to the role of the parametric dependence on $s$. From a Wilsonian point of view, the former corresponds to a choice of coarse graining scheme, while the latter corresponds to the coarse graining scale. We are primarily interested on the scale dependence of the theory, and therefore we take the point of view that a specific cutoff function has been chosen, and view \eqref{family} as a one-parameter family of theories.\footnote{In principle, physical quantities, such as critical exponents, are independent of the coarse graining scheme (see for example the universality of the one-loop beta function for marginal couplings, which we discuss in appendix \ref{app1}), but approximations generally spoil this property. Scheme dependence is thus an important issue, which has been greatly developed into the art of optimization \cite{Litim:2001up}, but we will not discuss it further here.}
We thus obtain a one-parameter family of free energies,
\begin{equation}\label{free}
W_{s}:=\log \mathcal{Z}_{s}[\bar{J},J],
\end{equation}
and by their Legendre transform, a one-parameter family of effective actions, collectively called \textit{effective average action}.
To be more precise, the effective average action $\Gamma_{s}$ is defined as:
\begin{equation}\label{legendre}
\Gamma_{s}[\bar{\phi},\phi]+\langle \bar{\phi}, R_{s}\phi\rangle=\langle \bar{J}, \phi\rangle+\langle \bar{\phi},J \rangle-W_{s}[\bar{J},J],
\end{equation}
where the source $J$ is to be expressed as a function of the effective mean field $\phi$ via the solution of
\begin{equation}\label{meanfield}
\phi =\dfrac{\delta W_{s}}{\delta \bar{J}}.
\end{equation}
The previous properties concerning the cutoff term $\Delta S_{s}$ mean, at the effective average action level, that:\\

$\bullet$ $\Gamma^{int}_{s=\ln \Lambda}=S_{int}$, so that when all the fluctuations are frozen, the effective average action coincides with the initial ``microscopic'' action.\\

$\bullet$ $\Gamma_{s=-\infty}=\Gamma$, meaning that when all the fluctuations are integrated out, the effective average action coincides with the full effective action. \\

\noindent
The definition \eqref{family} of the modified action can be usually interpreted as a modification of the propagator. For a standard Gaussian measure, with usual kinetic action, this property is obvious, but it is not so obvious for our model, for which the covariance has been defined via equation \eqref{propagator} rather than by an explicit kinetic action.\footnote{Alternatively, we can define our covariance $C$ in its momentum representation, directly via equation \eqref{p-propagator}. In this case, we see that because of the projector $\delta(\sum_i p_{i})$, the definition of the inverse $C^{-1}$, which usually appears in the kinetic action, is slightly subtle and we prefer to present a more formal derivation involving only $C$.} To prove this point, we use the Wick-theorem. 
Consider the following covariance:
\begin{equation}\label{propagatornew}
C_s(\vec{p},\vec{p}\,{}')=\dfrac{\delta\Big(\sum_ip_i\Big)}{\vec{p}\,{}^2+m^2+R_s(\vec{p})}\delta_{\vec{p},\vec{p}\,{}'},
\end{equation}
and the two Gaussian integrations:
\begin{align}
C_s(\vec{p},\vec{p}\,{}')&:=\int d\mu_{C_s}(T,\bar{T})\bar{T}(\vec{p})T(\vec{p}\,{}')\\
J_s(\vec{p},\vec{p}\,{}')&:=\int d\mu_{C}(T,\bar{T})e^{-\sum_{\vec{p}\,{}''}R_s\bar{T}(\vec{p}\,{}'')T(\vec{p}\,{}'')}\bar{T}(\vec{p})T(\vec{p}\,{}'),
\end{align}
where $d\mu_C$ is the normalized Gaussian integration: $\int d\mu_C=1$. In a first step, we will prove that $J_s\propto C_s$. This comes obviously from the Wick theorem. Using the derivative representation of a Gaussian integral:
\begin{align}
J_s(\vec{p},\vec{p}\,{}')&=\exp{\left( \dfrac{\delta}{\delta \psi}C\dfrac{\delta}{\delta \bar{\psi}}\right)}e^{-\sum_{\vec{p}}R_s\bar{T}(\vec{p})T(\vec{p})}\bar{T}(\vec{p})T(\vec{p}\,{}')\Big|_{T, \bar{T}=0}\\\nonumber
&=\dfrac{d}{dx}\exp{\left( \dfrac{\delta}{\delta \psi}C\dfrac{\delta}{\delta \bar{\psi}}\right)}e^{-\langle \bar{T},(R_s+xL) T\rangle}\Big|_{T, \bar{T},x=0}\\\nonumber
&=\dfrac{d}{dx}e^{-\Tr\ln(1+C(R_s+xL))}\Big|_{x=0}=C_s(\vec{p},\vec{p}\,{}')\times \det \Big[\dfrac{C_s}{C}\Big],
\end{align}
where the elements of the (super-) matrix $L$ are defined as $L_{\vec{p}_1,\vec{p}_2}=\delta_{\vec{p}_1,\vec{p}}\delta_{\vec{p}_2,\vec{p}\,{}'}$. Next, consider the following Gaussian integral:
\begin{equation}
J'_s:=\int d\mu_{C_s}(T,\bar{T})\prod_{j=1}^{N}\bar{T}^{j}(\vec{p}_{j})T^{j}(\vec{p}_{j}^{\,\prime}).
\end{equation}
Using the Wick theorem, and the previous result, we obtain:
\begin{align*}
J'_s&=\sum_{\mathcal{\pi}_N}\prod_{j}\left( \int d\mu_{C'}(T,\bar{T})\bar{T}^{\mathcal{\pi}_N(j)}(\vec{p}_{\mathcal{\pi}_N(j)})T^{j}(\vec{p}_{j}^{\,\prime})\right)\\
&=\sum_{\mathcal{\pi}_N}\prod_{j}\Bigg(\det\left[\dfrac{C}{C_s}\right]\int d\mu_{C}(T,\bar{T})e^{-\langle\bar{T},R_s T\rangle}\times{T}^{\mathcal{\pi}_N(j)}(\vec{p}_{\mathcal{\pi}_N(j)})T^{j}(\vec{p}_{j}^{\,\prime})\Bigg),
\end{align*}
where $\pi_N$ is the permutation group of $N$ elements. Now, using the Wick theorem, the big parenthesis can be written as:
\begin{align}\label{J}
\nonumber I_s&:=\det\left[\dfrac{C}{C_s}\right]\int d\mu_{C}(T,\bar{T})e^{-\langle\bar{T},R_s T\rangle}{T}^{\mathcal{\pi}_N(j)}(\vec{p}_{\mathcal{\pi}_N(j)})T^{j}(\vec{p}_{j}^{\,\prime})\\\nonumber
&=\sum_n \dfrac{(-1)^n}{n!}\det\left[\dfrac{C}{C_s}\right]\int d\mu_{C}(T,\bar{T})\times\langle\bar{T},R_s T\rangle^n \bar{T}^{\mathcal{\pi}_N(j)}(\vec{p}_{\mathcal{\pi}_N(j)})T^{j}(\vec{p}_{j}^{\,\prime})\\
&=\det\left[\dfrac{C}{C_s}\right]\sum_n \sum_{p}^{n}\dfrac{(-1)^{n-p+p}}{n!}\dfrac{n!}{p!(n-p)!}\\\nonumber
&\qquad \qquad\times\left\langle \langle\bar{T},R_s T\rangle^p\right\rangle_{C} \left\langle \langle\bar{T},R_s T\rangle^{n-p}\bar{T}^{\mathcal{\pi}_N(j)}(\vec{\theta}_{\mathcal{\pi}_N(j)})T^{j}(\vec{\theta}_{j}^{\,\prime})\right\rangle_{C,NV},
\end{align}
where the brackets $\langle.\rangle_C$ mean the Gaussian contractions with covariance $C$, and $\langle.\rangle_{C,NV}$ with subscript $NV$ select only the ``non-vacuum'' contractions (i.e. involving the external fields only). The Cauchy decomposition formula allows to rewrite the double sums as:
\begin{align*}
I_s^{j}&=\sum_p \left\langle \dfrac{(-1)^p}{p!} \langle\bar{T},R_s T\rangle^p\right\rangle_{C}\times\sum_n \left\langle \langle\bar{T},R_s T\rangle^{n}\bar{T}^{\mathcal{\pi}_N(j)}(\vec{p}_{\mathcal{\pi}_N(j)})T^{j}(\vec{p}_{j}^{\,\prime})\right\rangle_{C,N.V}\\
&=\left\langle e^{-\langle\bar{T},R_s T\rangle}\right\rangle_{C} \times \left\langle e^{-\langle\bar{T},R_s T\rangle}\bar{T}^{\mathcal{\pi}_N(j)}(\vec{p}_{\mathcal{\pi}_N(j)})T^{j}(\vec{p}_{j}^{\,\prime})\right\rangle_{C,N.V}.
\end{align*}
The first term $ \left\langle e^{-\int \bar{\psi}\Delta \psi}\right\rangle_{C}$ is nothing but the trivial Gaussian integral, i.e. the normalization of the Gaussian measure $d\mu_{C_s}$:
\begin{equation*}
\left\langle e^{-\langle\bar{T},R_s T\rangle}\right\rangle_{C}=e^{-\Tr\ln(1+C\Delta)}=\det(1+C\Delta)^{-1}=\det\left(\dfrac{C_s}{C}\right),
\end{equation*}
implying finally:
\begin{equation*}
I_s^j=\left\langle e^{-\langle\bar{T},R_s T\rangle}\bar{T}^{\mathcal{\pi}_N(j)}(\vec{p}_{\mathcal{\pi}_N(j)})T^{j}(\vec{p}_{j}^{\,\prime})\right\rangle_{C,N.V}
\end{equation*}
Reporting this result into the equation \eqref{J} for $J'_s$ leads to:
\begin{align}
J'_s=\sum_{\pi_N}\prod_{j}I_s^j=\det\left(\dfrac{C}{C_s}\right)\left\langle e^{-\langle\bar{T},R_s T\rangle}\right\rangle_{C}\sum_{\pi_N}\prod_{j}I_s^j.
\end{align}
Finally, because of the fact that the factor $\left\langle e^{-\langle\bar{T},R_s T\rangle}\right\rangle_{C}$ is nothing but the vacuum contribution of the measure $d\mu_Ce^{-\langle\bar{T},R_s T\rangle}$, we find:
\begin{align} \label{measure-s}
\int &d\mu_{C_s}(T,\bar{T})\prod_{j=1}^{N}\bar{T}^{j}(\vec{p}_{j})T^{j}(\vec{p}_{j}^{\,\prime})=\det\left(\dfrac{C}{C_s}\right)\int d\mu_{C}(T,\bar{T})e^{-\langle\bar{T},R_s T\rangle}\prod_{j=1}^{N}\bar{T}^{j}(\vec{p}_{j})T^{j}(\vec{p}_{j}^{\,\prime}).
\end{align}
Not surprisingly, this result is exactly what is expected for a standard Gaussian integration with well defined kinetic action. Hence, it follows from the properties of the Gaussian integration that the definition \eqref{family}, with modification of the action, can be interpreted as a modification of the covariance, i.e.
\begin{equation}\label{family2}
\mathcal{Z}'_{s}[\bar{J},J]:=\int d\mu_{C_s}(\bar{\psi},\psi)e^{-S_{int}(\bar{\psi},\psi)+\langle \bar{J},\psi \rangle+\langle\bar{\psi},J\rangle}.
\end{equation}
In fact, the flow equations for \eqref{family} and \eqref{family2} differ for vacuum terms due to the $\det(C_s)$ implicit in the measure of  \eqref{family2} (see \eqref{measure-s}), but as usual such vacuum terms are completely unimportant.\\

Because of the gauge invariance of our model, it is useful to define the projector into the gauge invariant field $\hat{P}$ as:
\begin{align}
\hat{P}:\mathcal{T}&\rightarrow \mathbb{G}\\\nonumber
\hat{P}:\psi\rightarrow\hat{P}[\psi](\theta_1,...,\theta_d)&=\int \dfrac{d\eta}{2\pi}\psi(\theta_1+\eta,...,\theta_d+\eta),
\end{align}
and the invariant field subspace as $\mathbb{G}=\ker(\hat{P}-\mathbb{I})\subset \mathcal{T}$. Interestingly, because of the gauge constraint in the propagator, the gauge constraint is implemented, at the graph level, at each colored line hooked to a black or white vertex. Hence, the mean field $\phi$, defined by \eqref{meanfield} lives in $\mathbb{G}$. The gauge symmetry is then dynamically implemented by the propagator, and contaminates each $N$-points functions : $G_N \in \otimes_{i=1}^{N}\mathbb{G}$.\\

For the 1-point function $\phi$, this property can be proved as follow. From the general expression \eqref{Schwinger}, $\phi$ can be expanded as:
\begin{equation}
\phi(\vec{\theta})=\sum_{\mathcal{G} \,connected}\dfrac{1}{s(\mathcal{G})}\left( \prod_{b\in\mathcal{G}}(-\lambda_b)\right)\mathcal{A}_{\mathcal{G}}(\vec{\theta}),
\end{equation}
where $\mathcal{A}_{\mathcal{G}}(\vec{\theta})$ is a Feynman amplitude of order $V(\mathcal{G})$ with one external point. A typical graph contributing to this expansion is depicted on the Figure \ref{fig9} below, and its explicit expression has the following structure:
\begin{align}
\mathcal{A}_{\mathcal{G}}(\vec{\theta})=\int \prod_{j=1}^4d\vec{\theta}_j\prod_{j'=2}^4d\vec{\theta}_{j'}&C_s(\vec{\theta},\vec{\theta}_1)\prod_{i=2}^4C_s(\vec{\theta}_i,\vec{\theta}_i')\mathcal{W}^{(1)}_{\vec{\theta}_1,\vec{\theta}_2,\vec{\theta}_3,\vec{\theta}_4}\bar{\mathcal{A}}_{\mathcal{G}}(\{\vec{\theta}_2',\vec{\theta}_3',\vec{\theta}_4'\}),
\end{align}
\noindent
where for convenience we have used of the propagator $C_s$ defined by \eqref{propagatornew} unlike of the propagator $C$. From its definition, the propagator lives in $\mathbb{G}\otimes \mathbb{G}$, because it verifies : $$\hat{P}C_s\hat{P}=C_s.$$\noindent Hence, it follows that the three vector indices $\vec{\theta}_2, \vec{\theta}_3, \vec{\theta}_4$ of the (super-)tensor $\mathcal{W}^{(1)}_{\vec{\theta}_1,\vec{\theta}_2,\vec{\theta}_3,\vec{\theta}_4}$ are projected into the gauge invariant space, and this tensor is equivalent to :
\begin{equation}
\bar{\mathcal{W}}^{(1)}_{\vec{\theta}_1,\vec{\theta}_2,\vec{\theta}_3,\vec{\theta}_4}=\int\prod_{i=2}^4\frac{d\beta_i}{2\pi}\mathcal{W}^{(1)}_{\vec{\theta}_1,\bar{\vec{\theta}}_2,\bar{\vec{\theta}}_3,\bar{\vec{\theta}}_4},
\end{equation}
where $\bar{\vec{\theta}}_i:=(\theta_{1i}+\beta_i,\theta_{2i}+\beta_i,...,\theta_{di}+\beta_i)$. The consequences are clearer in Fourier components:
\begin{align}
\bar{\mathcal{W}}^{(1)}_{\vec{\theta}_1,\vec{\theta}_2,\vec{\theta}_3,\vec{\theta}_4}&=\int\prod_{i=2}^4\frac{d\beta_i}{2\pi}\sum_{\vec{p}_i}\mathcal{W}^{(1)}_{\vec{p}_1,\vec{p}_2,\vec{p}_3,{\vec{p}}_4}e^{i\vec{p}_1\cdot \vec{\theta}_1}\times\prod_{j=2}^4e^{i\vec{p}_j\cdot \vec{\theta}_j}e^{i\beta_j\big(\sum_{k=1}^dp_{jk}\big)}\\\nonumber
&=\sum_{\vec{p}_i}\mathcal{W}^{(1)}_{\vec{p}_1,\vec{p}_2,\vec{p}_3,{\vec{p}}_4}\prod_{j=1}^4e^{i\vec{p}_j\cdot \vec{\theta}_j}\prod_{l=2}^4\delta\Bigg(\sum_{k=1}^dp_{lk}\Bigg).
\end{align}
The index structure of the bubble vertex implies that $p_{1k}=p_{2k}\,\forall k\neq 1$,$p_{21}=p_{31}$ and $p_{11}=p_{41}$. But, because of the Kronecker delta, $p_{41}=p_{31}$, implying $p_{11}=p_{21}$ and finally $\vec{p}_1=\vec{p}_2$. Hence $\vec{p}-1$ satisfies the closure constraint: $\sum_{k=1}^6p_{1k}=0$, and the last term of the previous equation can then be rewritten as:
\begin{align}
\nonumber\bar{\mathcal{W}}^{(1)}_{\vec{\theta}_1,\vec{\theta}_2,\vec{\theta}_3,\vec{\theta}_4}&=\sum_{\vec{p}_i}\mathcal{W}^{(1)}_{\vec{p}_1,\vec{p}_2,\vec{p}_3,{\vec{p}}_4}\prod_{j=1}^4e^{i\vec{p}_j\cdot \vec{\theta}_j}\delta\Bigg(\sum_{k=1}^dp_{jk}\Bigg)\\
&=\quad\int\prod_{i=1}^4\frac{d\beta_i}{2\pi}\mathcal{W}^{(1)}_{\bar{\vec{\theta}}_1,\bar{\vec{\theta}}_2,\bar{\vec{\theta}}_3,\bar{\vec{\theta}}_4}\in\otimes_{i=1}^4\mathbb{G},
\end{align}
meaning that $\hat{P}[\phi]=\phi$, or that $\phi\in \mathbb{G}$, thus proving our claim.

\bigskip
\begin{center}
\includegraphics[scale=1]{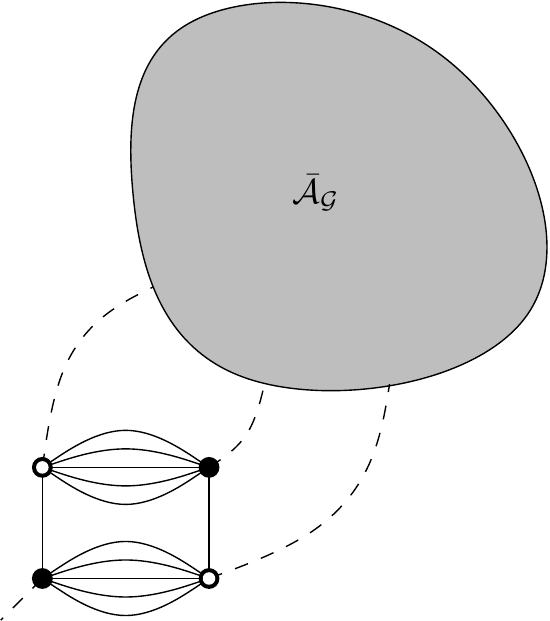}
\captionof{figure}{Typical graph contributing to $\phi$}\label{fig9}
\end{center}
\bigskip
\noindent
Using the Legendre transform \eqref{legendre} and the definition \eqref{family}, we obtain formally the so-called ``Wetterich equation'', describing the evolution of the effective average action. We have:

\begin{proposition}\emph{\textbf{(Wetterich equation)}}\\
For a given cutoff $R_s$, the effective average action $\Gamma_s$ satisfies the following partial differential equation:
\begin{align}\label{We}
\partial_s \Gamma_s=&\sum_{\vec{p}\in \mathbb{Z}^d} \partial_s R_s(\vec{p})\cdot \big[\Gamma_s^{(2)}+R_s\big]^{-1}(\vec{p},\vec{p})\,\delta\Bigg(\sum_{i=1}^d p_i\Bigg)\\\nonumber
=&\Tr\left[\hat{P}\dfrac{\partial_s R_s}{\Gamma_s^{(2)}+R_s}\right],
\end{align}
where $\Tr$ is the ''super-trace'', meaning the trace over the block-indices, and $\Gamma_s^{(2)} \equiv \tfrac{\delta^2\Gamma_s}{\delta \bar{\phi}\delta \phi}$.
The presence of $\hat{P}$ restricts the sums over the subspace of $\mathbb{Z}^6$ which we denote by $\mathcal{P}$ and define by : $\mathcal{P}=\{\vec{p}\in\mathbb{Z}^6|\sum_ip_i=0\}$. 
\end{proposition}
\textit{\textbf{Proof}:} The proof is rather standard \cite{Wetterich:1992yh,Morris:1993qb}, but it is useful to review it because of the TGFT context and the appearance of the projector $\hat{P}$ in the equation. Applying the operator $\partial_s$ on the two members of the equation \eqref{legendre}, we have
\begin{align}\label{eqref1}
\partial_s\Gamma_{s}&=\langle \partial_s\bar{J},\phi\rangle+\langle \bar{\phi},\partial_s J\rangle-\partial_sW_{s}
-\langle\partial_s\bar{J},\dfrac{\delta W_{s}}{\delta \bar{J}}\rangle - \langle\dfrac{\delta W_{s}}{\delta J},\partial_s J\rangle -\langle\bar{\phi},\partial_sR_{s}\phi\rangle.
\end{align}
The term $\partial_sW_{s}$ can be easily computed from the definitions \eqref{family} and \eqref{free}:
\begin{equation}\label{eqref2}
\partial_sW_{s} =-\sum_{\vec{p}\in \mathbb{Z}^d} \partial_s R_s(\vec{p})\Bigg(\dfrac{\delta^2W_{s}}{\delta J_{\vec{p}} \delta \bar{J}_{\vec{p}}}+\dfrac{\delta W_{s}}{\delta J_{\vec{p}}}\dfrac{\delta W_{s}}{\delta \bar{J}_{\vec{p}}}\Bigg).
\end{equation}
Using \eqref{meanfield}, many terms cancel in  \eqref{eqref1}, and we are left with
\begin{equation}\label{eqref3}
\partial_s\Gamma_{s} =\sum_{\vec{p}\in \mathbb{Z}^d} \partial_s R_s(\vec{p})\dfrac{\delta^2W_{s}}{\delta J_{\vec{p}} \delta \bar{J}_{\vec{p}}}.
\end{equation}

Deriving  \eqref{meanfield} with respect to $J$ we find 
\begin{equation}
 \frac{\delta^2W_{s}}{\delta J\delta \bar{J}} = \frac{\delta\phi}{\delta J} , 
\end{equation}
while deriving \eqref{legendre} with respect to $\phi$ and then  $\bar \phi$ gives 
\begin{equation}
\frac{\delta^2\Gamma_s}{\delta \bar{\phi}\delta \phi} = \frac{\delta J}{\delta\phi} - R_s.
\end{equation}
Therefore we obtain, in matrix notation:
\begin{equation}\label{invert}
\dfrac{\delta^2W_{s}}{\delta J\delta \bar{J}}\Bigg[\dfrac{\delta^2\Gamma_s}{\delta \bar{\phi}\delta \phi}+R_s\Bigg]=\mathbb{I}.
\end{equation}
The second functional derivative $W_s^{(2)}$ is the $2$-points function which, as the effective mean field $\phi=W_{s}^{(1)}$, is gauge invariant: $W_{s}^{(2)}\in\mathbb{G}\otimes{\mathbb{G}}$. Hence, the previous equation \eqref{invert} means that $\Gamma^{(2)}+R_s$ is the inverse of $\mathcal{W}^{(2)}$ on the subspace $\mathcal{P}$ only. 
Finally, using \eqref{invert} in \eqref{eqref3}, and the fact that, because $\mathcal{W}^{(2)}$ is a gauge invariant matrix : $\mathcal{W}^{(2)}\in \mathbb{G}\otimes \mathbb{G} \rightarrow \hat{P}\mathcal{W}^{(2)}\hat{P}=\mathcal{W}^{(2)}$, we find the Wetterich equation \eqref{We}. 
\begin{flushright}
$\square$
\end{flushright}

\subsection{The truncation procedure}

One of the many advantages of the FRG is the possibility to make an approximation directly at the level of the effective average action. The latter is in principle a complicated functional, containing infinitely many terms, and therefore an exact solution of the Wetterich equation is generally beyond reach. One of the most used approximation methods, besides perturbation theory, consists in truncating the space of functionals in which $\Gamma_s$ is defined, retaining only a finite-dimensional subspace, or a simple infinite-dimensional one.
This allows to obtain a closed and tractable system of differential equations, which can be analytically studied in some cases, or numerically integrated for most of the cases. 
As an organizational principle for such truncations, one usually starts off by Taylor expanding the effective average action  in powers of $\vec{p}$. Truncating such expansion to some given order is expected to be a valid approximation in a context where we are interested in the long distance physics. 
Subsequently, one can proceed in expanding each order in powers of the fields, and truncate that expansion too, thus leading to a finite-dimensional subspace of functionals.
The main idea is that by systematically expanding such subspace one should observe some convergence pattern in the values of physical quantities, such as the critical exponents, and thus justify a posteriori the approximation procedure. A neat example of this method is provided by the Ising universality class, associated to the Wilson-Fisher fixed point in three dimensions \cite{Canet:2003qd}. However, when applying the FRG in a new context, before reaching such advanced level of systematization, it is often necessary to start from the simplest possible approximation in such class of truncations, to work out the formalism and understand possible outcomes. 
In such case, the most usual choice essentially boils down to writing $\Gamma_s$ exactly in the same form as the bare action we have in mind, but with coupling, mass and wave function normalization depending on $s$. The latter is the approximation we will adopt here. \\

More precisely, we will henceforth consider  the following simple truncation:
\begin{align}\label{ansatz}
\Gamma_s[\phi,\bar{\phi}]&=\int d\vec{\theta} \bar{\phi}(\vec{\theta})\big[-Z_s\Delta_{\vec{\theta}}+m_s^2\big]\phi(\vec{\theta})\\\nonumber
&\quad +\lambda_s\sum_i \int \left(\prod_{j=1}^{4} d\vec{\theta}_j\right)\mathcal{W}_{\vec{\theta}_1,\vec{\theta}_2,\vec{\theta}_3,\vec{\theta}_4}^{(i)} \phi(\vec{\theta}_1)\bar{\phi}(\vec{\theta}_2)\phi(\vec{\theta}_3)\bar{\phi}(\vec{\theta}_4),
\end{align}
where $\Delta_{\vec{\theta}}$ is the Laplacian operator on $U(1)^{\otimes d}$, $\lambda_s$ and $m_s$ are the effective coupling and mass at the scale $s$, and $Z_s$ is the field strength normalization. \\

Evaluating the second derivative of \eqref{ansatz}, we get, in momentum representation:
\begin{align}\label{secmom}
\Gamma_s^{(2)}(\vec{p},\vec{p}^{\,\prime})&=\Big[Z_s\vec{p}\,{}^2+m_s^2\Big]\hat{P}_{\vec{p},\vec{p}^{\,\prime}}+2\lambda_s \sum_{\{\vec{p}_i\}}\hat{P}_{\vec{p},\vec{p}_1}\hat{P}_{\vec{p}^{\,\prime},\vec{p}_2}\sym\mathcal{W}_{\vec{p}_1,\vec{p}_2,\vec{p}_3,\vec{p}_4}T_{\vec{p}_3}\bar{T}_{\vec{p}_4},
\end{align}
where $\mathcal{W}^{(i)}_{\vec{p}_1,\vec{p}_2,\vec{p}_3,\vec{p}_4}$ is given in \eqref{W_p}, and we have introduced the notation $\mathcal{W}=\sum_{i=1}^d \mathcal{W}^{(i)}$,
\begin{equation}
\sym\mathcal{W}_{\vec{p}_1,\vec{p}_2,\vec{p}_3,\vec{p}_4}:=\mathcal{W}_{\vec{p}_1,\vec{p}_2,\vec{p}_3,\vec{p}_4}+\mathcal{W}_{\vec{p}_3,\vec{p}_2,\vec{p}_1,\vec{p}_4}.
\end{equation}
and
\begin{equation}
\hat{P}_{\vec{p}_1,\vec{p}_2}=\delta_{\vec{p}_1,\vec{p}_2}\delta\Bigg(\sum_{i=1}^dp_i\Bigg).
\end{equation}
Let us also define:
\begin{equation}
F_s({\vec{p},\vec{p}^{\,\prime}}) := 2 \sum_{\{\vec{p}_i\}}\hat{P}_{\vec{p},\vec{p}_1}\hat{P}_{\vec{p}^{\,\prime},\vec{p}_2}\sym\mathcal{W}_{\vec{p}_1,\vec{p}_2,\vec{p}_3,\vec{p}_4}T_{\vec{p}_3}\bar{T}_{\vec{p}_4}
\end{equation}
and
\begin{equation}
K_s({\vec{p}}):=\Big(Z_s\vec{p}\,{}^2+m_s^2+R_s(\vec{p})\Big).
\end{equation}
The next step is the choice of the regulator $R_s$, for which we adopt Litim's cutoff  \cite{Litim:2001up}:
\begin{equation}\label{mass}
R_s(\vec{p})=Z_s (e^{2s}-\vec{p}\,{}^2)\Theta(e^{2s}-\vec{p}\,{}^2),
\end{equation}
where $\Theta$ stands for the Heaviside step function. The special dependence on $\vec{p}\,{}^2$, and the inclusion of a wave function renormalization, will lead to substantial simplifications.
Applying the derivative operator on $R_s$, we find:
\begin{equation}\label{derivativer}
\partial_s R_s(\vec{p})=\Big\{\partial_s Z_s(e^{2s}-\vec{p}\,{}^2)+2Z_se^{2s}\Big\}\Theta(e^{2s}-\vec{p}\,{}^2).
\end{equation}
Using the equations \eqref{secmom} and \eqref{derivativer} in the flow equation \eqref{We}, we will obtain the renormalization group flow equations, which are the subject of the next section. 
\bigskip
\subsection{Renormalization group flow equations}	
\label{sectiongfe}
Using our ansatz \eqref{ansatz} in the Wetterich equation \eqref{We}, expanding its r.h.s. of in power of $\phi$, and truncating at order $\phi^4$, we find:
\begin{align*}
r.h.s.=&\sum_{\vec{p}\in\mathcal{P}}\dfrac{\partial_sR_s}{K_s(\vec{p})}\Bigg[1-\lambda_s \frac{F_s(\vec{p},\vec{p})}{K_s(\vec{p})}+\lambda_s^2\sum_{\vec{p}^{\,\prime}\in \mathcal{P}}\dfrac{F_s(\vec{p},\vec{p}^{\,\prime})F_s(\vec{p}^{\,\prime},\vec{p})}{K_s(\vec{p})K_s(\vec{p}^{\,\prime})}+\mathcal{O}(\phi^6)\Bigg]. 
\end{align*}
The RG flow equations for coupling, mass and wave function renormalization can be obtained from this expansion, identifying the corresponding powers of $\bar{\phi}\phi$. \\

$\bullet$ \textbf{Equations for $m_s$ and $Z_s$}. Identifying the terms of order $\bar{\phi}\phi$ leads to:
\begin{align}
\nonumber \sum_{\vec{p}\in\mathcal{P}}(\partial_s m_s^2+\partial_s Z_s \vec{p}\,{}^2) T_{\vec{p}}\bar{T}_{\vec{p}}&=-2\lambda_s\sum\limits_{\substack{\vec{p}\in\mathcal{P}\\|\vec{p}|\leq e^s}}\dfrac{\big(\partial_s Z_s(e^{2s}-\vec{p}\,{}^2)+2Z_se^{2s}\big)}{(Z_se^{2s}+m_s^2)^2}\\\label{e1}
&\times \sum_{\vec{q}_1,\vec{q}_2}\sym\mathcal{W}_{\vec{p},\vec{p},\vec{q}_1,\vec{q}_2}T_{\vec{q}_1}\bar{T}_{\vec{q}_2}.
\end{align}
We start the analysis with some considerations about the r.h.s.. Because of the form of the interaction matrices $\mathcal{W}^{(i)}$ and $T_{\vec{p}} \in \mathbb{G}$:
\begin{align} \label{TT1}
\sum_{\vec{p}\in \mathcal{P}}\sum_{\vec{q}_1,\vec{q}_2}f(\vec{p}\,{}^2)\mathcal{W}^{(i)}_{\vec{p},\vec{p},\vec{q}_1,\vec{q}_2}T_{\vec{q}_1}\bar{T}_{\vec{q}_2}=\sum_{\vec{p},\vec{p}^{\,'}\in \mathcal{P}}f(\vec{p}\,{}^2)T_{\vec{p}^{\,'}}\bar{T}_{\vec{p}^{\,'}}\delta_{p_i,p'_i}.
\end{align}
Defining $\vec{p}_{\bot}$ to be the set $(p_1,\ldots,p_{i-1},p_{i+1},\ldots,p_d)$, and
\begin{align}
\tilde{f}(p_i^2) &= \sum_{\vec{p}_{\bot}}f(\vec{p}\,{}^2)\, \delta\left(\sum_{j=1}^d p_j\right),
\end{align}
equation \eqref{TT1} can be rewritten as:
\begin{align}
\sum_{\vec{p}\in \mathcal{P}}\sum_{\vec{q}_1,\vec{q}_2}f(\vec{p}\,{}^2)\mathcal{W}^{(i)}_{\vec{p},\vec{p},\vec{q}_1,\vec{q}_2}T_{\vec{q}_1}\bar{T}_{\vec{q}_2}=\sum_{\vec{p}\in \mathcal{P}}\tilde{f}(p_i^2) \,  T_{\vec{p}}\bar{T}_{\vec{p}}.
\end{align}
In the same way, because of the closure constraint, 
\begin{align} \label{TT2}
\sum_{\vec{p}\in \mathcal{P}}\sum_{\vec{q}_1,\vec{q}_2}f(\vec{p}\,{}^2)\mathcal{W}^{(i)}_{\vec{q}_1,\vec{p},\vec{p},\vec{q}_2}T_{\vec{q}_1}\bar{T}_{\vec{q}_2}=\sum_{\vec{p}\in \mathcal{P}} f(\vec{p}\,{}^2)T_{\vec{p}}\bar{T}_{\vec{p}}.
\end{align}
%
We can therefore rewrite \eqref{e1} as
\begin{align} \label{e1bis}
\sum_{\vec{p}\in\mathcal{P}}(\partial_s m_s^2+\partial_s Z_s \vec{p}\,{}^2) T_{\vec{p}}\bar{T}_{\vec{p}}
&=-2\lambda_s\sum_{\vec{p}\in\mathcal{P}} \left(\sum_{i=1}^d \tilde{f}(p_i^2) + d\, f(\vec{p}\,{}^2) \right)  T_{\vec{p}}\bar{T}_{\vec{p}} ,
\end{align}
where
\begin{align}
f(\vec{p}\,{}^2) &= \left(A+B \,\vec{p}\,{}^2\right) \theta(e^{2s}-\vec{p}\,{}^2),
\end{align}
and we have defined
\begin{align} \label{eq:A}
A &:= \dfrac{\big(\partial_s Z_s+2Z_s\big)e^{2s}}{(Z_se^{2s}+m_s^2)^2},\\ \label{eq:B}
B &:= - \dfrac{\partial_s Z_s}{(Z_se^{2s}+m_s^2)^2}.
\end{align}

We want to satisfy \eqref{e1bis} up to order $\vec{p}\,{}^2$, meaning that we want to expand 
\begin{align}
\sum_{i=1}^d \tilde{f}(p_i^2) + d\, f(\vec{p}\,{}^2) = d \,\left(\tilde{f}(0) +  f(0) \right) +\vec{p}\,{}^2\, \left( \tilde{f}'(0) + d\, f'(0)\right) + O(\vec{p}\,{}^4),
\end{align}
equate the coefficients of $\vec{p}\,{}^0$ and $\vec{p}\,{}^2$, 
\begin{align} \label{proj_m}
\partial_s m_s^2 &= -2\lambda_s\,d \,\left(\tilde{f}(0) +  f(0) \right) ,\\  \label{proj_Z}
\partial_s Z_s &= -2\lambda_s\, \left( \tilde{f}'(0) + d\, f'(0)\right)
\end{align}
and discard the rest. The expansion of $f(\vec{p}\,{}^2)$ is trivial, as the step function is equal to one at all orders of the Taylor expansion, and we thus have $f(0) = A$ and $f'(0)=B$. The expansion of $\tilde{f}(p_i^2)$ is instead slightly more involved. We can first write $\tilde{f}(p_i^2)$ as
\begin{equation}
\tilde{f}(p_i^2) = A S_1(p_i^2) + B S_2(p_i^2),
\end{equation}
where
\begin{align} \label{eq:S_1}
S_1(k^2)&:= \sum_{\vec{p}_{\bot}} \delta\Bigg(k+\sum_{l\neq i}p_l\Bigg) \theta(e^{2s}-k^2-\vec{p}_{\bot}\,{}^2),\\  \label{eq:S_2}
S_2(k^2)&:= \sum_{\vec{p}_{\bot}} (k^2+\vec{p}\,{}^2_{\perp})\delta\Bigg(k+\sum_{l\neq i}p_l\Bigg) \theta(e^{2s}-k^2-\vec{p}_{\bot}\,{}^2),
\end{align}
so that
\begin{align} \label{proj_m_1}
\partial_s m_s^2 &= -2\lambda_s\,d \,\left(A\,(1+S_1(0))+B\,S_2(0) \right) ,\\  \label{proj_Z_1}
\partial_s Z_s &= -2\lambda_s\, \left( A\,S'_1(0)+ B\,(d+S'_2(0))\right).
\end{align}
Since we will be mostly interested in the large-$s$ limit (the small-$s$ case will be treated separately), we can approximate the sums by integrals, replacing the Kronecker deltas by Dirac deltas.
The support of the integrals is in the intersection of the hyperplane of equation $k+ \sum_{l\neq i}p_l=0$ and the $(d-1)$-ball of radius $\sqrt{e^{2s}-k^2}$. Note that the Kronecker delta of the closure constraint can be rewritten as $k+\vec{p}_{\bot}\cdot \vec{n}=0$, where $\vec{n}=(1,1,...1)$ is the vector with all components equals to $1$ in $\mathbb{R}^{d-1}$. Using the rotational invariance of our integral, we can chose one of our coordinate axis to be in the direction $\vec{n}$. If we choose the axis $2$ in this direction, our constraint writes as $\delta(k+ p'_2|\vec{n}|)=\delta( k+p'_2\sqrt{d-1})=\delta( k/\sqrt{d-1}+ p'_2)/\sqrt{d-1}$, and we find the following integral approximation:
\begin{align}
S_1(k^2)&\simeq\frac{1}{\sqrt{d-1}}\Omega_{d-2}\bigg[e^{2s}-\frac{dk^2}{d-1}\bigg]^{\frac{d-2}{2}} =: \tilde S_1(k^2)\\
S_2(k^2)&\simeq\frac{1}{\sqrt{d-1}}\Big[\frac{dk^2}{d-1}+\frac{d-2}{d}\bigg(e^{2s}-\frac{dk^2}{d-1}\bigg)\Big]\Omega_{d-2}\bigg(e^{2s}-\frac{dk^2}{d-1}\bigg)^{\frac{d-2}{2}} =: \tilde S_2(k^2),
\end{align}
where $\Omega_{d}:=\pi^{d/2}/\Gamma(d/2+1)$ is the volume of the unit $d$-ball.
The integral approximations are easily expanded to yield:
\begin{align}  \label{eq:I_1}
I_1 &:= \tilde S_1(0)=\frac{1}{\sqrt{d-1}}\Omega_{d-2}e^{(d-2)s},\\  \label{eq:I_2}
I_2 &:= \tilde S_2(0) =\frac{1}{\sqrt{d-1}}\dfrac{d-2}{d}\Omega_{d-2}e^{ds},\\  \label{eq:I_3}
I_3 &:=  \tilde S'_1(0) =-\frac{1}{\sqrt{d-1}}\dfrac{d (d-2)}{2 (d-1)}\Omega_{d-2}e^{(d-4)s},\\  \label{eq:I_4}
I_4 &:=  \tilde S'_2(0)=-\frac{1}{\sqrt{d-1}}\dfrac{d (d-4)}{2 (d-1)}\Omega_{d-2}e^{(d-2)s}.
\end{align}
Therefore we obtain
\begin{align} \label{proj_m_2}
\partial_s m_s^2 &= -2\lambda_s\,d \,\left(A\,(1+I_1)+B\,I_2 \right) ,\\  \label{proj_Z_2}
\partial_s Z_s &= -2\lambda_s\, \left( A\,I_3+ B\,(d+I_4)\right),
\end{align}
which, using \eqref{eq:A}-\eqref{eq:B} and \eqref{eq:I_1}-\eqref{eq:I_4}, translate into
\begin{empheq}[box=\fbox]{equation}\label{flow1}
\partial_s m_s^2=-2\lambda_sZ_s\dfrac{\Big(2\frac{\Omega_{d-2}}{\sqrt{d-1}}e^{ds}+d e^{2s}\Big)\eta_s+2d\Big(\frac{\Omega_{d-2}}{\sqrt{d-1}}e^{ds}+ e^{2s}\Big)}{(Z_se^{2s}+m_s^2)^2},
\end{empheq}
and
\begin{empheq}[box=\fbox]{equation}\label{flow2}
 \eta_s=\frac{2d(d-2)}{(d-1)^{3/2}}\dfrac{\lambda_s\Omega_{d-2}e^{(d-2)s}}{(Z_se^{2s}+m_s^2)^2-2\lambda_s\Big[\frac{d}{(d-1)^{3/2}}\Omega_{d-2} e^{(d-2)s}+d\Big]},
\end{empheq}
where we have defined 
\begin{equation}
\eta_s:=\frac{\partial}{\partial s}\log(Z_s).
\end{equation}

$\bullet$ \textbf{Equation for $\lambda_s$}. In order to obtain the equation describing the running coupling constant flow, we must identify the correct melonic structure among the terms of order $(\bar{\phi}\phi)^2$ in the r.h.s. of the flow equation, i.e. in
\begin{align*}
\sum_{\vec{p}\in\mathcal{P}}\dfrac{\partial_sR_s}{K_s(\vec{p})}\lambda_s^2\sum_{\vec{p}^{\,\prime}\in \mathcal{P}}\dfrac{F_s(\vec{p},\vec{p}^{\,\prime})F_s(\vec{p}^{\,\prime},\vec{p})}{K_s(\vec{p})K_s(\vec{p}^{\,\prime})}.
\end{align*}
The expression above involves a term of the form $$\sym\mathcal{W}_{\vec{p},\vec{p}\,{}',\vec{p}_1,\vec{p}_2}\sym\mathcal{W}_{\vec{p},\vec{p}\,{}',\vec{p}_1',\vec{p}_2'},$$ 
\noindent
giving three types of contributions, $$\mathcal{W}^{i}_{\vec{p},\vec{p}\,{}',\vec{p}_1,\vec{p}_2}\mathcal{W}^{j}_{\vec{p},\vec{p}\,{}',\vec{p}_1',\vec{p}_2'},$$ $$\mathcal{W}^{i}_{\vec{p_1},\vec{p}\,{}',\vec{p},\vec{p_2}}\mathcal{W}^{j}_{\vec{p},\vec{p}\,{}',\vec{p}_1',\vec{p}_2'},$$ $$\mathcal{W}^{i}_{\vec{p}_1,\vec{p}\,{}',\vec{p},\vec{p}_2}\mathcal{W}^{j}_{\vec{p}_1',\vec{p}\,{}',\vec{p},\vec{p}_2'},$$ all depicted in figure \ref{fig4} below. \\

The connectivity of the original melonic interaction induces a selection rule for the interactions generated by the r.h.s of the flow equation. This selection rule is based on the study of the ``ultra-local'' version of the interactions depicted in figures \ref{fig4}a, \ref{fig4}b and \ref{fig4}c. \\

\begin{center}
\includegraphics[scale=1]{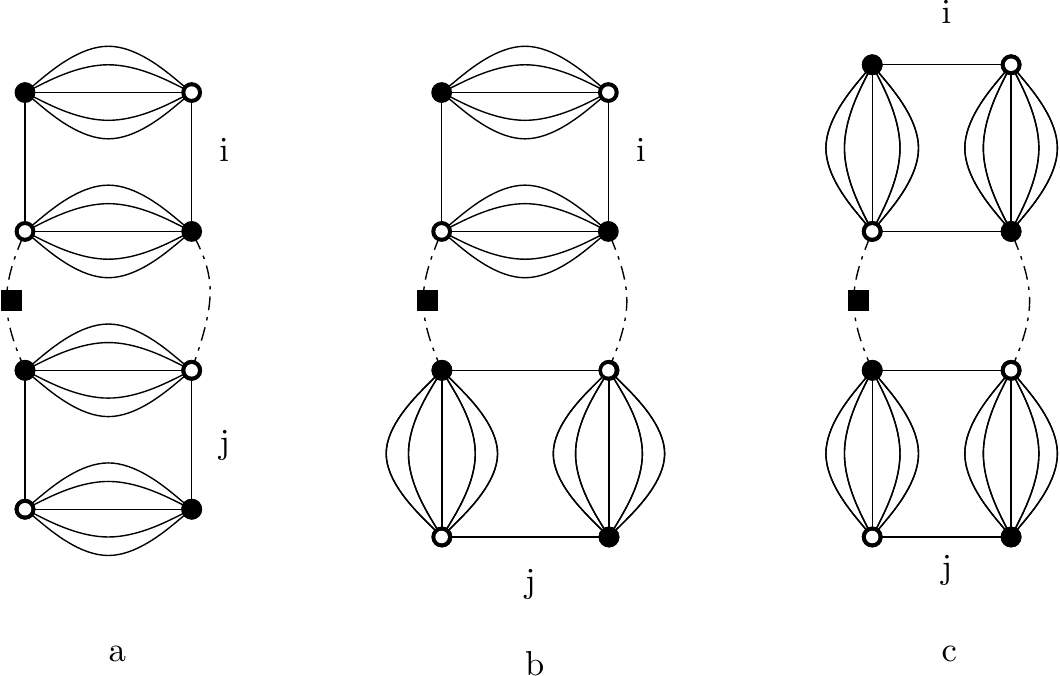} 
\captionof{figure}{Interactions involved in the r.h.s of the Wetterich equation. The dash-dotted line represents a contraction involving the operator $K_s^{-1}$, and the dash-dotted line with a square represents a contraction involving the operator $\partial_sR_s\,K_s^{-2}$}\label{fig4}
\end{center}
In order to define this selection rule, we need the help of the following definition, illustrated in figure \ref{fig5}:\\

\begin{definition}\label{def1}
\textbf{(contraction operation)}. Let $b$ an interaction bubble with $L(b)$ dash-dotted lines. Let $L_0=\{l_i\}\subset L(b)$ an ordered subset of dash-dotted lines in $b$. The contracted graph $b/L_0$ is obtain from $b$ by:\\


\noindent
$\bullet$ Deleting the line $l_i\in L_0$ and its two (black and white) end vertices and all the colored lines joining these two vertices.\\
$\bullet$ Identifying the colored line linked to the deleted black vertex with the line of corresponding color linked to the white vertex. \\
$\bullet$ Repeat for $l_{i+1}$. 
\end{definition}
\medskip
\begin{center}
\includegraphics[scale=1]{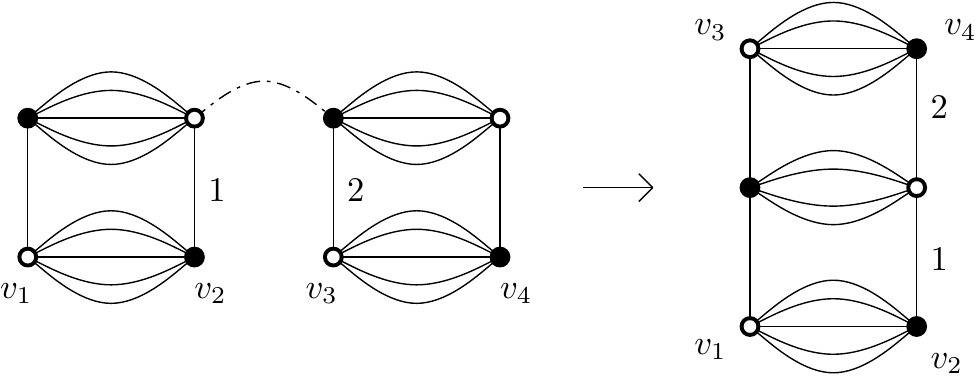} \\
\captionof{figure}{Contraction of a dotted line between two vertices. Some black and white vertices are labeled as $v_i$ in order to facilitate the understanding of the picture.}\label{fig5}
\end{center}
The purpose of this contraction procedure is just to allow us to obtain the connectivity structure of an interaction, and therefore we do not keep track of the black square (the insertion of the cutoff operator) in the definition above and in the figures. Of course we will keep track of that once we have identified the correct structures. 
As an example of contraction, consider the interaction depicted in \ref{fig4}a, for which we obtain, after contraction over the two dash-dotted lines, the result depicted in figure \ref{fig6}.
\begin{center}
\includegraphics[scale=1]{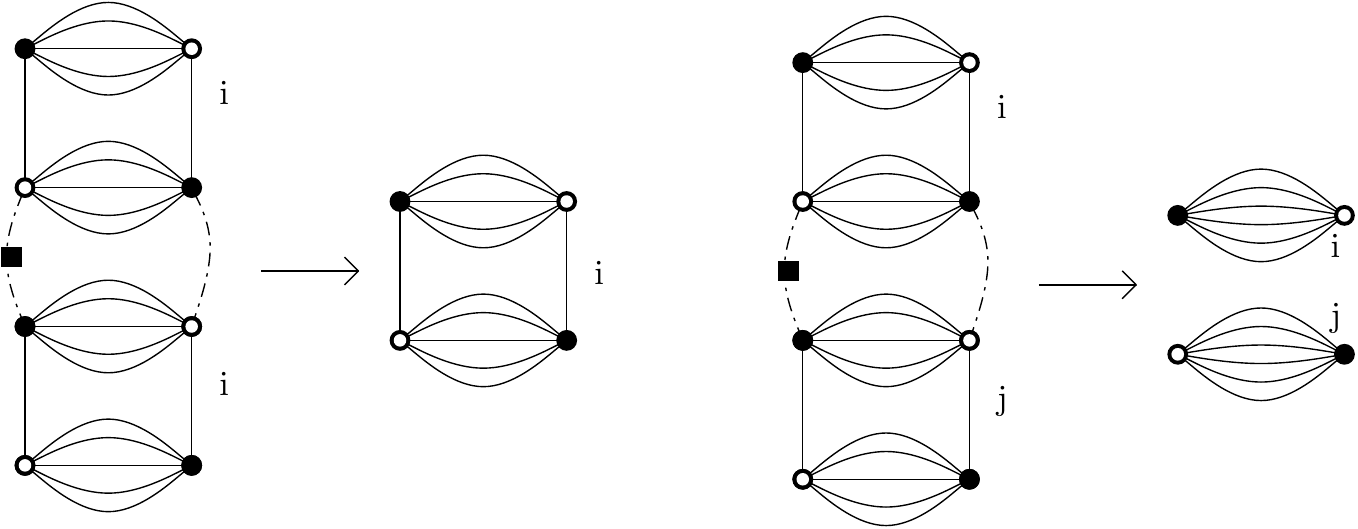} \\
\captionof{figure}{Connecting structure of the interactions \ref{fig4}a.}\label{fig6}
\end{center}
We observe that the first contraction gives exactly a vertex of the original form, while the second contraction gives a disconnected vertex corresponding to the square of a mass-term, which is outside of our truncation, and thus will be discarded. As a second example, the case $i=1$ and $j=2$ for the last \ref{fig4}c case gives, after contraction of its two dash-dotted lines,  the result depicted in figure \ref{fig7}, showing that the interaction does not have the connectivity structure of a melonic bubble, and will also be discarded.
\begin{center}
\includegraphics[scale=1]{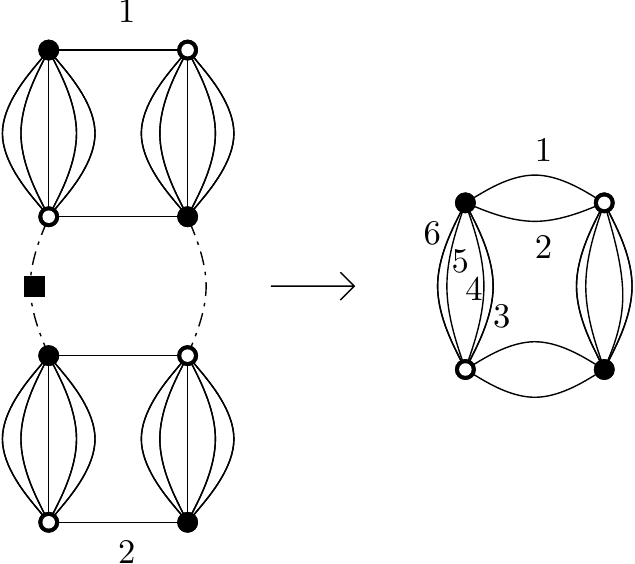} \\
\captionof{figure}{Connecting structure of the interactions \ref{fig4}c with $i\neq j$}\label{fig7}
\end{center}

Of all the possibles interactions, only those of type \ref{fig4}a and \ref{fig4}c with $i=j$, and \ref{fig4}b with $i\neq j$, respect the connectivity of the original melonic interaction.
Therefore, we retain only such terms  for the computation of $\partial_s\lambda_s$, i.e. we make the approximation
\begin{align}\label{approx1}
\sym\mathcal{W}_{\vec{p},\vec{p}\,{}',\vec{p}_1,\vec{p}_2}\sym\mathcal{W}_{\vec{p}\,{}',\vec{p},\vec{p}_3,\vec{p}_4} 
&\simeq\sum_{i=1}^{d}\Bigg[\mathcal{W}^{(i)}_{\vec{p},\vec{p}\,{}',\vec{p}_1,\vec{p}_2}\mathcal{W}^{(i)}_{\vec{p}\,{}',\vec{p},\vec{p}_3,\vec{p}_4}
+\mathcal{W}^{(i)}_{\vec{p}_1,\vec{p}\,{}',\vec{p},\vec{p}_2}\mathcal{W}^{(i)}_{\vec{p}_3,\vec{p},\vec{p}\,{}',\vec{p}_4}\\ \nonumber
&\quad\qquad +  \mathcal{W}^{(i)}_{\vec{p},\vec{p}\,{}',\vec{p}_1,\vec{p}_2} \sum_{j\neq i} \mathcal{W}^{(j)}_{\vec{p}_3,\vec{p},\vec{p}\,{}',\vec{p}_4}
\Bigg]\\ \nonumber
&=:  \left(\bar{\sym}^2\mathcal{W}\right)_{\vec{p},\vec{p}\,{}',\vec{p}_1,\vec{p}_2,\vec{p}_3,\vec{p}_4},
\end{align}
after which we obtain the equation
\begin{align}
\partial_s\lambda_s\, \mathcal{W}_{\vec{p}_1,\vec{p}_2,\vec{p}_3,\vec{p}_4}T_{\vec{p}_1}\bar{T}_{\vec{p}_2}T_{\vec{p}_3}\bar{T}_{\vec{p}_4} = & 4\lambda_s^2\sum_{\vec{p},\vec{p}\,{}'\in\mathcal{P}}
\dfrac{\big(\partial_s Z_s(e^{2s}-\vec{p}\,{}^2)+2Z_se^{2s}\big)}{(Z_se^{2s}+m_s^2)^3} \theta\left(e^{2s}-\vec{p}\,{}^2\right)\\ \nonumber
& \times \left(\bar{\sym}^2\mathcal{W}\right)_{\vec{p},\vec{p}\,{}',\vec{p}_1,\vec{p}_2,\vec{p}_3,\vec{p}_4} T_{\vec{p}_1}\bar{T}_{\vec{p}_2}T_{\vec{p}_3}\bar{T}_{\vec{p}_4} .
\end{align}

As before, we want to satisfy this equation to the desired order in $\vec{p}_i{}^2$, in this case meaning to zeroth order.
The easiest way to accomplish that is to project the equation onto the field $T_{\vec{p}}^{(0)}=\prod_{i=1}^d \delta_{p_i}^{0}$, leading to the equation:
\begin{align*}
d\partial_s\lambda_s = 4\lambda_s^2&\sum_{\vec{p},\,|\vec{p}|\leq e^s}\sum_{\vec{p}\,{}'}\dfrac{\big(\partial_s Z_s(e^{2s}-\vec{p}\,{}^2)+2Z_se^{2s}\big)}{(Z_se^{2s}+m_s^2)^3}\times\delta\Bigg(\sum_{i=1}^d p_i\Bigg)
\left(\bar{\sym}^2\mathcal{W}\right)_{\vec{p},\vec{p}\,{}',\vec{0},\vec{0},\vec{0},\vec{0}} .
\end{align*}
Using \eqref{W_p}, and taking into account  the closure constraint $\sum_i p_i=0$, we find $\mathcal{W}^{(i)}_{\vec{0},\vec{p},\vec{p}\,{}',\vec{0}}=\prod_{i=1}^d \delta_{p_i}^{0}\prod_{j=1}^d \delta_{p_j,p'_j}$ and $\mathcal{W}^{(i)}_{\vec{p},\vec{p}\,{}',\vec{0},\vec{0}}=\delta_{p_i,0}\delta\big(\sum_{j\neq i}p_j\big) \prod_{j=1}^d \delta_{p_j,p'_j}$, and as a consequence:
\begin{align}
\delta\Bigg(\sum_{i=1}^d p_i\Bigg)\left(\bar{\sym}^2\mathcal{W}\right)_{\vec{p},\vec{p}\,{}',\vec{0},\vec{0},\vec{0},\vec{0}} 
=\left(d^2\prod_{i=1}^d\delta_{p_i}^0 
+\sum_{i=1}^d\delta_{p_i}^0 \delta\left(\sum_{j\neq i}p_j\right)\right)\delta_{\vec{p},\vec{p}\,{}'}.
\end{align}
As for the mass and anomalous dimension, the flow of $\lambda_s$ can be expressed in term of $S_1$ and $S_2$, defined in \eqref{eq:S_1} and \eqref{eq:S_2}, giving
\begin{equation} \label{proj_lambda}
\partial_s\lambda_s=4\lambda_s^2\dfrac{\Big(de^{2s}+e^{2s}S_1(0)-S_2(0)\Big)\partial_sZ_s+2e^{2s}\Big(d+S_1(0)\Big)Z_s}{(Z_se^{2s}+m_s^2)^3}.
\end{equation}
or, in term of the integral approximations \eqref{eq:I_1} and \eqref{eq:I_2}: 
\begin{empheq}[box=\fbox]{align}\label{flow3}
\partial_s\lambda_s=4\lambda_s^2\dfrac{\Big(de^{2s}+\frac{2}{d}\frac{\Omega_{d-2}}{\sqrt{d-1}}e^{ds}\Big)\partial_sZ_s+2\Big(de^{2s}+\frac{\Omega_{d-2}}{\sqrt{d-1}}e^{ds}\Big)Z_s}{(Z_se^{2s}+m_s^2)^3}.
\end{empheq}
\subsection{RG equations for ``dimensionless'' parameters}
The RG flow equation obtained above describes the evolution of the couplings in our truncation with respect to the ``scale-time'' $s$. But, as in the quantum field theory or condensed matter models, a trivial contribution to the evolution comes from the ``dimension'' of the operator associated to these couplings. 
We define the dimensionless parameters $\bar{\lambda}_s$ and $\bar{m}_s$ as
\begin{align}
\lambda_s=e^{(6-d)s}\bar{\lambda}_s \qquad m_s=e^s\bar{m}_s,
\end{align}
using the results about the dimension of the ``local operators'' introduced in section $1.2$. 
The flow equations for the  dimensionless couplings are easily deduce from the previous ones. The equation for $\eta_s$ is unchanged, and the equations for $\lambda_s$ and $m_s$ become:
\begin{align*}
\partial_s& \bar{m}_s^2=-2\bar{m}_s^2-2\bar{\lambda}_s\dfrac{\Big(2\frac{\Omega_{d-2}}{\sqrt{d-1}}+d e^{(2-d)s}\Big)\partial_s Z_s+2d\Big(\frac{\Omega_{d-2}}{\sqrt{d-1}}+ e^{(2-d)s}\Big)Z_s}{(Z_s+\bar{m}_s^2)^2}
\end{align*}
\begin{align*}
\partial_s&\bar{\lambda}_s=(d-6)\bar{\lambda}_s+4\bar{\lambda}_s^2\dfrac{\Big(d\,e^{(2-d)s}+\frac{2}{d}\frac{\Omega_{d-2}}{\sqrt{d-1}}\Big)\partial_sZ_s+2\Big(d\,e^{(2-d)s}+\frac{\Omega_{d-2}}{\sqrt{d-1}}\Big)Z_s}{(Z_s+\bar{m}_s^2)^3}.
\end{align*}

Another more physical redefinition concerns the relations between the different renormalization parameters. The couplings $\lambda_s$ and $m_s$ which appear in the original model are not the effective mass and effective coupling at the scale $s$. Indeed, the evolution of the wave function renormalization $Z_s$ from $1$ to an arbitrary value, modifies the effective couplings, which become : $\bar{\lambda}_s=Z_s^2\bar{\lambda}_s^r$, $\bar{m}_s=\sqrt{Z}_s\bar{m}_s^r$. Hence, from the two previous equations, we finally deduce:
\begin{align}\label{flowbis1}
\partial_s& \bar{m}_s^{r\,2}=-2\Big(1+\frac{\eta_s}{2}\Big)\bar{m}_s^{r\,2}-2\bar{\lambda}_s^r\dfrac{\Big(2\frac{\Omega_{d-2}}{\sqrt{d-1}}+d e^{(2-d)s}\Big)\eta_s+2d\Big(\frac{\Omega_{d-2}}{\sqrt{d-1}}+ e^{(2-d)s}\Big)}{(1+\bar{m}_s^{r\,2})^2},
\end{align}
\begin{align}\label{flowbis2}
\partial_s&\bar{\lambda}_s^r=(d-6-2\eta_s)\bar{\lambda}_s^r+4\bar{\lambda}_s^{r\,2}\dfrac{\Big(d\,e^{(2-d)s}+\frac{2}{d}\frac{\Omega_{d-2}}{\sqrt{d-1}}\Big)\eta_s+2\Big(d\,e^{(2-d)s}+\frac{\Omega_{d-2}}{\sqrt{d-1}}\Big)}{(1+\bar{m}_s^{r\,2})^3},
\end{align}
with:
\begin{equation}\label{flowbis3}
\eta_s=\frac{2d(d-2)}{(d-1)^{3/2}}\dfrac{\Omega_{d-2}\bar{\lambda}_s^r }{(1+\bar{m}_s^{r\,2})^2-2\bar{\lambda}_s^r\Big[\frac{d}{(d-1)^{3/2}}\Omega_{d-2}+de^{(2-d)s}\Big]}.
\end{equation}
Note that, similarly to what was found in \cite{BBGO}, the system of differential equations we just obtained is not autonomous, in the sense that even after switching to dimensionless couplings it maintains an explicit dependence on the RG scale $s$.
This was interpreted in \cite{BBGO} (based on more general cases in which a similar phenomenon takes place, e.g.  \cite{Benedetti:2014gja}) as consequence of the implicit scale set by the compactness of the group, i.e. the radius of $S^1\simeq U(1)$, an interpretation corroborated by the recent results of \cite{Geloun:2015qfa}, where autonomous equations are obtained in the non-compact limit $U(1)\to \mathbb{R}$.

\section{Large-$s$ approximation and fixed points}
\label{Sec:large-s}
In this section we examine the flow equation in the large-$s$ approximation, or in the UV limit. In this sector, the equations are much simpler, since they become autonomous, and thus it is possible to find fixed points, and their relevant or irrelevant perturbations. 

\subsection{Large $s$ approximation, vicinity of the Gaussian fixed point}
\label{sectiongaussian}
In the UV regime, corresponding to the large-$s$ limit, the previous equations \eqref{flowbis1}, \eqref{flowbis2}, and \eqref{flowbis3}, for $d>2$, reduce to:
\begin{align}\label{flowbis12}
\partial_s \bar{m}_s^{r\,2}=-2\Big(1+\frac{\eta_s}{2}\Big)\bar{m}_s^{r\,2}-2\bar{\lambda}_s^r\frac{\Omega_{d-2}}{\sqrt{d-1}}\dfrac{2\eta_s+2d}{(1+\bar{m}_s^{r\,2})^2},
\end{align}
\begin{align}\label{flowbis22}
\partial_s\bar{\lambda}_s^r=(d-6-2\eta_s)\bar{\lambda}_s^r+4\bar{\lambda}_s^{r\,2}\frac{\Omega_{d-2}}{\sqrt{d-1}}\dfrac{\frac{2}{d}\eta_s+2}{(1+\bar{m}_s^{r\,2})^3},
\end{align}
\begin{equation}\label{flowbis32}
\eta_s=\dfrac{2d(d-2)\Omega_{d-2}\bar{\lambda}_s^r }{(d-1)^{3/2}(1+\bar{m}_s^{r\,2})^2-2d\Omega_{d-2}\bar{\lambda}_s^r}.
\end{equation}
These equation form an autonomous system, whose fixed points can be studied with standard methods. A trivial fixed point occurs for $m_s^r=\lambda_s^r=0$, corresponding to the so called ``Gaussian fixed point'' (GFP). In order to study the stability of this fixed point, we expand the previous RGEs around this fixed point, at second order in the coupling constant. We obtain the following system (we limit the development of $\eta_s$ to the first order in $\lambda_s$, because the anomalous dimension appears always as a quantum correction):
\begin{equation}\label{betam}
\partial_s \bar{m}_s^{r\,2}=-2\bar{m}_s^{r\,2}+2d\frac{\Omega_{d-2}}{\sqrt{d-1}}\frac{3d-2}{(d-1)}\bar{\lambda}_s^r\bar{m}_s^{r\,2}-4d\frac{\Omega_{d-2}}{\sqrt{d-1}}\bar{\lambda}_s^r-\frac{8d(d-2)}{(d-1)^2}\Omega_{d-2}^2\bar{\lambda}_s^{r\,2}
\end{equation}
\begin{equation}\label{beta1}
\partial_s\bar{\lambda}_s^r=(d-6)\bar{\lambda}_s^r-4\bigg[\frac{d(d-2)}{d-1}-2\bigg]\frac{\Omega_{d-2}}{\sqrt{d-1}}\bar{\lambda}_s^{r\,2},
\end{equation}
\begin{equation}
\eta_s=\frac{2d(d-2)}{(d-1)^{3/2}}\Omega_{d-2}\bar{\lambda}_s^r.
\end{equation}
At $d=6$ the coupling $\bar{\lambda}_s^{r}$ become marginal, and using $\Omega_4=\pi^2/2$ we obtain:
\begin{equation}
\partial_s \bar{m}_s^{r\,2}=-2\bar{m}_s^{r\,2}+\frac{96\pi^2}{5\sqrt{5}}\bar{\lambda}_s^r\bar{m}_s^{r\,2}-\frac{12}{\sqrt{5}}\pi^2\bar{\lambda}_s^r-\frac{48}{5\sqrt{5}}\pi^2\bar{\lambda}_s^{r\,2}
\end{equation}
\begin{equation}\label{beta}
\partial_s\bar{\lambda}_s^r=-\frac{28\pi^2}{5\sqrt{5}}\bar{\lambda}_s^{r\,2},
\end{equation}
\begin{equation}
\eta_s=\frac{24\pi^2}{5\sqrt{5}}\bar{\lambda}_s^r.
\end{equation}
The minus sign in front of the r.h.s of equation \eqref{beta} means that the theory is asymptotically free. This result was found for several TGFT models \cite{BBGO,Rivasseau-AF,TGFTrenorm-Joseph,TGFTrenorm-others}, including our model \cite{Lahoche:2015ola,Dine}, and it seems what it is a generic property of TGFTs models. Appendix \ref{app2} give the one loop computation of the beta function, that we find in accordance with \ref{beta}, and with the well known universality of the one-loop beta function (see Appendix \ref{app1}). \\

For $d>6$, the theory is non-renormalizable, but the equations \eqref{betam} and \eqref{beta1} show the existence of a non-trivial fixed point for $\bar{\lambda}_s$ and $\bar{m}$ at:
\begin{equation}
\bar{\lambda}^*=\frac{1}{4}\frac{(d-6)(d-1)^{3/2}}{(d^2-4d+2)\Omega_{d-2}},
\end{equation}
\begin{equation}
\bar{m}^{*2}=\frac{d(d-6)(d-1)}{d^2-4d+2}\;\frac{3d^2-16d+16}{d(d-6)(3d-2)-4(d^2-4d+2)}.
\end{equation}
For $d=6$ the fixed point merges with the GFP, and for small $\epsilon=d-6$ it is at a small value of the couplings, and therefore the use of the expansion  \eqref{betam} and \eqref{beta1} is justified (but note that the mass grows rapidly and diverges at $d\simeq 6.65$).
In order to analyze the stability of this fixed point, we compute the first derivative, at the point $(\bar{m}^{2\,*},\bar{\lambda}^*)\equiv(g^*_1,g^*_2)$ of the functions $\beta_{i}$ defined by \eqref{betam} and \eqref{beta1}, where the index $i$ labels the couplings  $(\bar{m}^{r\,2},\bar{\lambda}^{r})\equiv(g_1,g_2)$. These derivatives build the ``stability matrix'' $\beta_{ij}^*=\partial\beta_i(g^*)/\partial g_j$, whose expression for general $d$ is not particularly enlightening. More important are its eigenvalues (the so-called ``critical exponents''), which, at leading order in $\epsilon$, are $-2+24\epsilon/7$ and $-\epsilon$, respectively, meaning that we have two relevant perturbations (an operator is said to be relevant if it becomes important in the IR limit, irrelevant if it disappears in the IR limit, and marginal for vanishing critical exponent). Therefore, for small $\epsilon>0$ we have asymptotic safety. Whether this survives at $\epsilon=1$ is a question that we will not address here. 

For $d=6$,  \eqref{betam} and \eqref{beta1} admit only the GFP, and the stability matrix reads:
\begin{equation}\beta^{*\,diag}_{ij}:=\begin{pmatrix}
-2&-\frac{12\pi^2}{\sqrt{5}}\\
0&0
\end{pmatrix},
\end{equation}
with eigenvalues $ (-2,0)\,$ (for the GFP the critical exponents coincide with the canonical dimension of the associated operators), and eigenvectors,
\begin{equation} 
h_1=\begin{pmatrix}
1 \\
0
\end{pmatrix} \qquad h_2=\begin{pmatrix}
-\frac{6\pi^2}{\sqrt{5}} \\
1
\end{pmatrix}.
\end{equation} 
The first direction is associated to a relevant operator (the mass), and the second one is marginal. To decide whether the marginal coupling is marginally relevant (i.e. asymptotically free) or marginally irrelevant (i.e. trivial) one has to go to next order in the expansion, as we have done in \eqref{beta}.

Rather than expanding \eqref{betam} and \eqref{beta1} in powers of the couplings, the flow can be studied numerically, and this is the subject of the next section (for the case $d=6$). 
\medskip

\subsection{Non-Gaussian fixed point with d=6}

In $d=6$, the autonomous system \eqref{flowbis12}, \eqref{flowbis22}, \eqref{flowbis32} has other non trivial fixed points than the trivial GFP. After a simple analysis of the equations, we find two fixed points, for the values:

\begin{equation}\label{FP}
\{\bar{\lambda}^*_{\pm},\bar{m}^*_{\pm}\}=\bigg\{\frac{\sqrt{5}(43309\mp 79\sqrt{1141})}{2135484\pi^2},\frac{-175\pm\sqrt{1141}}{234}\bigg\}.
\end{equation}
As for the Gaussian fixed point, we can study the stability of these fixed point, by computing the eigenvalues of the stability matrix, i.e. the critical exponents.
Diagonalizing the stability matrix at the two fixed points, we find:
\begin{equation}
\beta^{(+)\,diag}_{ij}\approx\begin{pmatrix}
4.86&0\\
0&-0.9
\end{pmatrix}
\end{equation}

\begin{equation}
\beta^{(-)\,diag}_{ij}\approx\begin{pmatrix}
262.8&0\\
0&7
\end{pmatrix},
\end{equation}
showing that the fixed point $\{\bar{\lambda}_{*\,+},\bar{m}_{*\,+}\}$ has one relevant and one irrelevant directions, whereas the other one has only irrelevant directions. 

Our result is very similar to the one obtained in \cite{BBGO} for a TGFT of rank-3 without closure constraint. As for that model, we find the existence of non-trivial fixed points (one of which carrying one relevant eigen-direction) in the critical dimension at which the model is asymptotically free. Note that, as we have seen in section \ref{sectiongaussian}, moving infinitesimally away from the critical dimension one loses asymptotic freedom, but a new non-trivial fixed point emerges from the Gaussian one. Such merging or splitting of fixed points near the critical dimension is a standard phenomenon, but it has nothing to do with the fixed points \eqref{FP}, which appear precisely at the critical dimension. Contrary to that, this is not a standard phenomenon, and it has to be attributed to the non-local structure of the TGFT interaction.

The numerical integration of the large-$s$ flow leads to the phase diagram of figure \ref{fig10}, which shows again a strong resemblance to the one found in \cite{BBGO}. Note that, because of our choice of truncation \eqref{ansatz} and parametrization, and in particular because of the choice of the regulator \eqref{mass}, the flow equations \eqref{flowbis12}, \eqref{flowbis22}, and \eqref{flowbis32} have singularities at $\bar{\lambda}^r=5^{3/2}(1+\bar{m}^{r\,2})^2/6\pi^2$.  A similar singularity was found in  \cite{BBGO}, where its origin is discussed in more detail. As a consequence of such singular line, the fixed point $\{\bar{\lambda}_{*\,-},\bar{m}_{*\,-}\}$ is disconnected from $\{\bar{\lambda}_{*\,+},\bar{m}_{*\,+}\}$ and the GFP.

The non-Gaussian fixed point $\{\bar{\lambda}_{*\,+},\bar{m}_{*\,+}\}$ is reminiscent of the Wilson-Fisher fixed point. It appears as an IR fixed point associated to a broken phase, with positive coupling and negative mass. More precisely, the region above the irrelevant trajectory at the NGFP can be interpreted as a symmetric phase, with coupling and mass of the same sign, and the region below this irrelevant trajectory as a broken phase, with negative mass and positive coupling. 

Within the present truncation, our conclusion about the existence of such non-trivial fixed point is only valid in the large cutoff limit, i.e. in the UV limit. Therefore, the occurrence of a non-trivial behavior associated to the broken phase, rests upon the survival of such fixed point in the IR, and this is why we will study the counterpart of the regime studied in this section, the small cutoff limit, in the next section. 

\begin{center}
\includegraphics[scale=0.8]{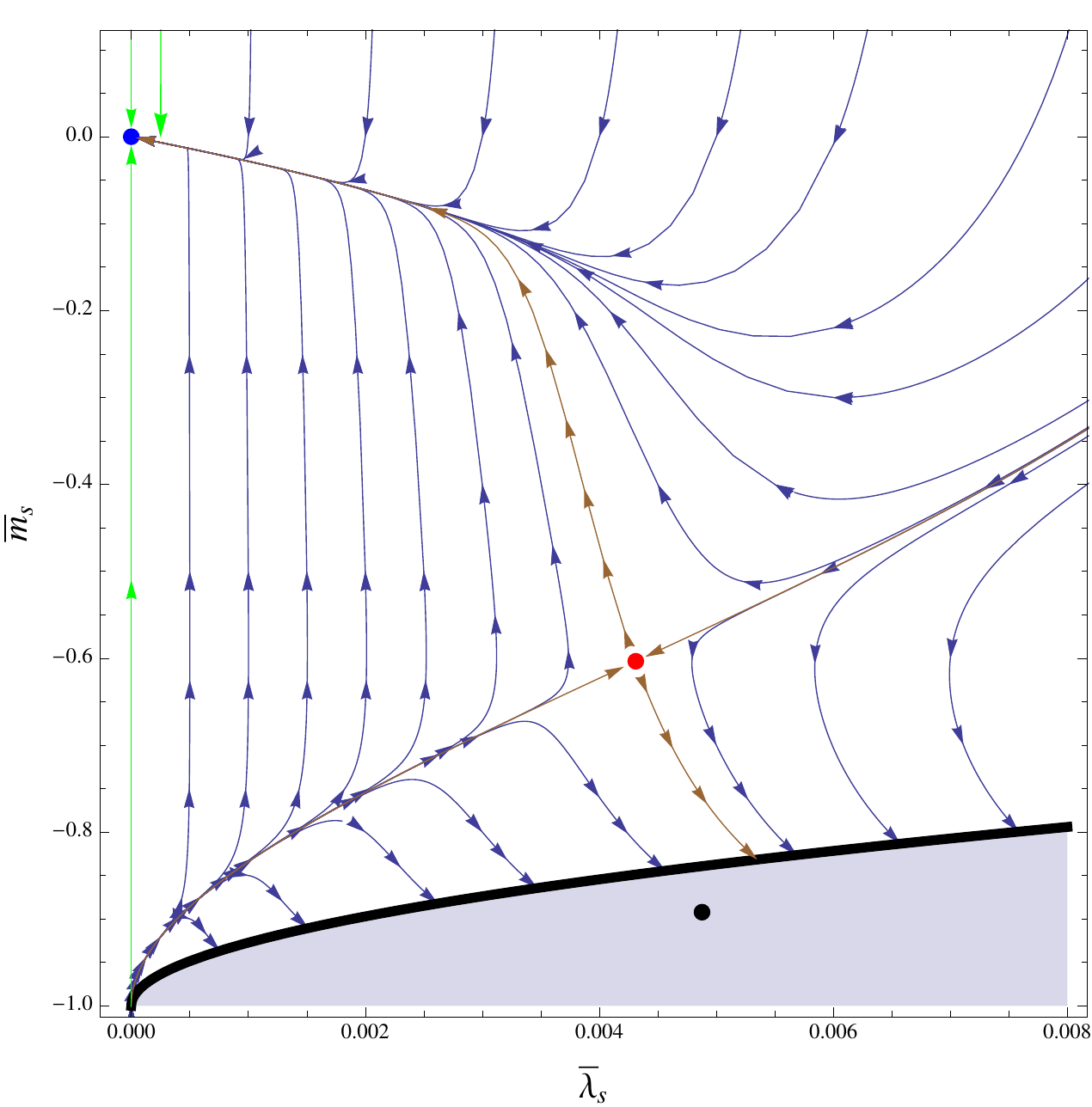} 
\captionof{figure}{The flow diagram at large cutoff. The blue dot is the GFP, while the red and black are the non trivial ones, labelled with $+$ and $-$, respectively. The black line corresponds to the singularity of the flow equations, because of which the shaded region is disconnected from the GFP. In green and brown are respectively the eigen-perturbation for the GFP and the NGFP.}\label{fig10}
\end{center}

\section{Small-$s$ limit and comments}
\label{Sec:small-s}
When the cutoff becomes very small, $e^s<1$, all the sums obtained in \ref{sectiongfe} reduce to a single element, the zero mode $\vec{p}=\vec{0}$. In this limit, for the sums defined in \eqref{eq:S_1} and \eqref{eq:S_2}, we find $S_1=1$ and $S_2=0$, independently of $k<e^s$.
Therefore, from equations \eqref{proj_m_1}, \eqref{proj_Z_1}, and \eqref{proj_lambda}, we find that the anomalous dimension vanishes, implying that we are free to choose $Z_s=1$, and that the beta functions becomes, in $d=6$:
\begin{equation}\label{betamlarge}
\partial_sm_s^2=-48\lambda_s\dfrac{e^{2s}}{(e^{2s}+m_s^2)^2},
\end{equation}
\begin{equation}
\partial_s\lambda_s=56\lambda_s^2\dfrac{e^{2s}}{(e^{2s}+m_s^2)^3}.
\end{equation}
Using the rescaling $m_s=e^s\bar{m}_s$ and $\lambda_s=e^{4s}\bar{\lambda}_s$ leads to the following autonomous system:
\begin{equation}
\partial_s\bar{m}_s^2=-2\bar{m}_s^2-48\dfrac{\bar{\lambda}_s}{(1+\bar{m}_s^2)^2}
\end{equation}
\begin{equation}
\partial_s\bar{\lambda}_s=-4\bar{\lambda}_s+56\dfrac{\bar{\lambda}_s^2}{(1+\bar{m}_s^2)^3}.
\end{equation}
In addition, as in the previous case, the system admits again a GFP, and a NGFP for the values:
\begin{equation}
\bar{\lambda}^*=\frac{49}{13718} \qquad \qquad \bar{m}^*=-\frac{12}{19}.
\end{equation}
In order to study the stability of these fixed points, we compute the matrix $\beta_{ij}$ as before. For the Gaussian fixed point, we find:
\begin{equation}
\beta^{GFP}_{ij}=\begin{pmatrix}
-2&-48\\
0&-4
\end{pmatrix},
\end{equation}
whose eigenvalues are $(-4,-2)$, corresponding to two relevant eigen-directions. 

For the non-Gaussian fixed point, we find the matrix:
\begin{equation}
\beta^{NG}_{ij}=\begin{pmatrix}
\frac{34}{7}&-\frac{17328}{49}\\
-\frac{42}{361}&4
\end{pmatrix},
\end{equation}
with eigenvalues $(76/7,-2)$, giving one relevant and one irrelevant directions. Interestingly, the critical exponents are integers, a property found also in  \cite{BBGO}, and which can traced back to the zero-dimensional nature of the theory in this regime (see discussion in \cite{BBGO}). As for the large $s$ limit case, the numerical integration of these flow equations give the Figure \ref{fig11} below.

Note that in this IR regime the rescaling for mass and coupling does not correspond to the canonical one defined in section \ref{sectiondim}. Again, this is a consequence of the fact that at small-$s$ our model is essentially a zero dimensional field theory, and our rescaling matches exactly with the one expected for a such theory, by fixing to zero the dimension of the (effective) action, and $1$ the dimension of the mass parameter. Indeed, by direct inspection of our truncation \eqref{ansatz}, we find $[\lambda]=4$. Because no phase transition is expected for such model, we have found an essential indication for the symmetric phase restoration in our model, as it should be on a compact space.\footnote{For a related discussion on effective dimensional reduction and symmetry restoration see also \cite{Guilleux:2015pma}.}
In the physics of phase transition, it is well known that phase transitions occur in the ``thermodynamic'' limit or, in other worlds, for a non-compact manifold. This limit, in our case, can be approached by restoring in our expressions the radius $L$ of $S^1\simeq U(1)$, here always fixed to 1, and by sending this radius to infinity. 
Such non-compact limit has been recently studied in \cite{Geloun:2015qfa}, confirming the fact that in such limit one is left with a picture identical to the large-$s$ regime, and thus with a non-trivial phase transition.

We have not extended our results in the intermediate cutoff regime, between small and large cutoff, but being the two asymptotic regimes qualitatively very similar to the ones in \cite{BBGO}, we expect the intermediate regime to add no new qualitative insight with respect to the analysis performed there.

\begin{center}
\includegraphics[scale=0.8]{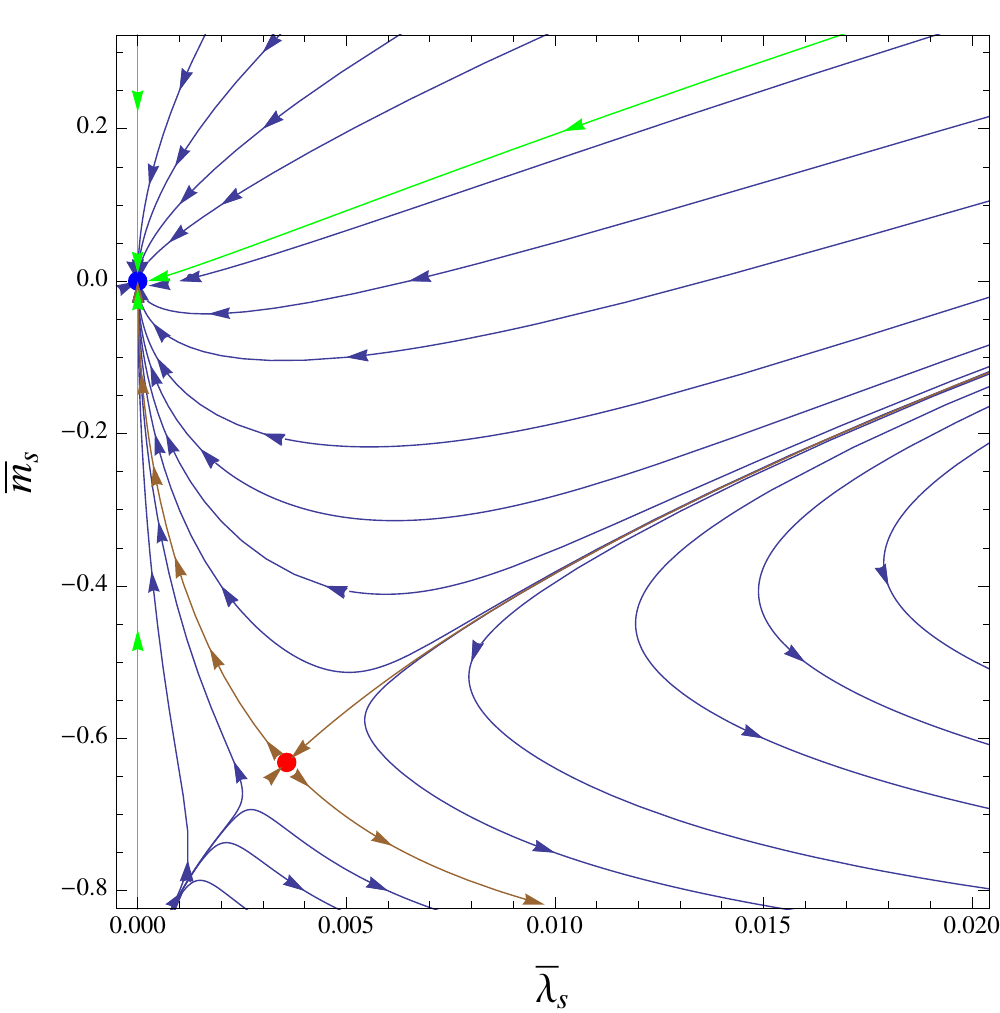} 
\captionof{figure}{The flow diagram at small cutoff. The color rules are the same as in figure \ref{fig10}. Ordinary trajectories are in blue, eigen-perturbations of the GFP are in green, and eigen-perturbation of the NGFP are in brown.}\label{fig11}
\end{center}

\section{Conclusion}
\label{Sec:concl}
In this paper we have developed the FRG formalism for a just-renormalizable model of TGFT on the compact group $U(1)^{\otimes 6}$, and we explored two regimes of the flow, for large and small cutoff, which are more suitable to analytical and numerical study. 

The lessons of our analysis can be summarized as follows. First, we have shown that the presence of the closure constraint can be incorporated in the FRG formalism without too much pain. We have thus obtained a Wetterich equation for this class of TGFTs.

Second, from our analysis of a specific truncation, we have recovered in the UV limit the asymptotic freedom found perturbatively in \cite{Lahoche:2015ola,Dine} for this model, and more generally observed for several TGFTs models \cite{TGFTrenorm-Joseph,TGFTrenorm-others,Rivasseau-AF,BBGO}. Furthermore, in the same limit, we have found a non-Gaussian fixed point, of which we have studied the stability, and identified the critical directions. Interestingly, this non-trivial fixed point has qualitatively similar characteristics to the Wilson-Fisher one, and it is associated to a phase transition in the flow diagram. 

Lastly, in the IR sector, due to the compactness of the group manifold, we obtained an effectively zero-dimensional theory, and thus we have symmetry restoration in the deep IR. This is not a surprise, given the compact nature of the manifold. The compactness of space can be viewed as an additional regularization, to be removed in order to study the non-compact limit \cite{Geloun:2015qfa}, or otherwise one can view the latter as the UV approximation, as we did here. In the non-compact limit one is left with just the large-$s$ regime of our analysis (because $s=\ln(kL)$, if $L$ is the radius of $S^1$ and $k$ the dimensionful RG scale), and thus with a non-trivial IR fixed point associated to a phase transition between a broken and a symmetric phase.

Overall, the above picture is strikingly similar to the one obtained in \cite{BBGO}, and in part to the one in  \cite{Geloun:2015qfa}, despite the models having essential differences.
The model in \cite{BBGO} is in three dimensions, whereas our is in six; the model in \cite{BBGO} has no closure constraint, whereas our does; and the model in  \cite{BBGO} has a kinetic term linear in momentum, whereas our is quadratic. These are all important differences, and the fact that the final result is nevertheless so similar points towards the greater importance that lies within the main similarity between the two models: the non-local melonic interaction.
We are therefore tempted to conjecture that such TGFTs with melonic interactions enjoy in quite some generality not only asymptotic freedom, but also a Wilson-Fisher-like fixed point. Of course, the approximation we employed here is the simplest one, and making our result more solid will require studying larger truncations. And more models will need to be studied in order to understand whether our conjecture applies to all asymptotically free TGFTs or only to some specific sub-class. Grasping also at a more qualitative level how such feature arises (e.g. similarly to how asymptotic freedom is understood to arise from the wave function renormalization taking place already at one loop \cite{Rivasseau-AF}) is also an open question. We are just at the beginning of the FRG investigations of TGFTs and hopefully these and other aspects will be clarified very soon.

\section*{Acknowledgments}

We would like to express our gratitude to Joseph Ben Geloun, Daniele Oriti, and Vincent Rivasseau, for many useful remarks during the preparation of this work, and on the final draft.

\newpage
\appendix


\section{Universality of the one-loop beta function for $\lambda$}
\label{app1}
In this section we review a well-known argument explaining why the one-loop beta function does not depend on the choice of the regularization, and we give an explicit example of this result for the computation of the beta function with FRG method around the Gaussian fixed point using a different regulator than the one used in the main part of the paper.

\subsection{A standard argument for one-loop universality}

Different computational scheme result in physical predictions being made in terms of different couplings.
Suppose that in some scheme we have a dimensionless coupling $\lambda$, whose flow equation reads
\begin{equation}
\partial_s \lambda = b_1 \lambda^2 + b_2 \lambda^3 + b_3 \lambda^4 + \ldots \,.
\end{equation}
In a different scheme we have a different coupling $\lambda'$ with flow
\begin{equation}
\partial_s \lambda' = b'_1 \lambda'^2 + b'_2 \lambda'^3 + b'_3 \lambda'^4 + \ldots \,.
\end{equation}
At tree level the two couplings are the same, because tree level does not involve any ambiguity, thus the relation between the two couplings must be of the form
\begin{equation} \label{lambdaprime}
\lambda' = \lambda+ C_1 \lambda^2 + C_2 \lambda^3  + \ldots \,.
\end{equation}
Now write
\begin{align}\nonumber
\partial_s \lambda' &= \frac{\partial \lambda'}{\partial\lambda} \partial_t \lambda\\ 
&= (1+2C_1\lambda+3 C_2\lambda^2+\ldots)(b_1 \lambda^2 + b_2 \lambda^3 + b_3 \lambda^4 + \ldots)\\ \nonumber
&= b_1 \lambda^2 + (2C_1 b_1+b_2) \lambda^3 + (3C_2 b_1 + 2 C_1 b_2 +b_3) \lambda^4 + \ldots\\ \nonumber
&= b_1 \lambda'^2 + b_2 \lambda'^3 + (b_3 -2 C_1 b_2 - C_1^2 b_1 +C_2 b_1) \lambda'^4 + \ldots \,,
\end{align}
where in the last row we used the inverse of \eqref{lambdaprime},
\begin{equation}
\lambda = \lambda'- C_1 \lambda'^2 - (C_2-2C_1^2) \lambda'^3  + \ldots,
\end{equation}
and we assumed that $\partial_t C_1 = 0$.
Therefore, under these assumptions, $b_1$ and $b_2$, i.e. the one- and two-loop coefficients of the beta function, are scheme independent.

Note that if $\partial_s C_1 \neq 0$, the flow of $\lambda'$ is not autonomous, unless $C_1=a s +b$, or unless the $s$-dependence of $C_1$ is related to the beta function of another coupling.
This is preciesely what happens in general within the FRG, which is a mass-dependent renormalization scheme \cite{Codello:2013bra}. However, when expanding the beta functions around zero mass, the above argument must hold true also for the FRG.

\subsection{Example with a different regulator}
As we mentioned in section \ref{Sec:FRG}, the choice of a cutoff function corresponds to a choice of coarse graining scheme. Different choices thus affect the flow, but we expect that physical quantities, such as the critical exponents, will not depend on it.
As we argued above, one such quantity is the one-loop coefficient of the beta function for the marginal coupling.
To illustrate this point more explicitly, we choose here a different regulator:
\begin{equation}
R_s^{\,\prime}(\vec{p})=\dfrac{Z_s\vec{p}\,{}^2+m_s^2}{\exp\Big[\frac{Z_s\vec{p}\,{}^2+m_s^2}{Z_se^{2s}}\Big]-1}\label{regulatorbis},
\end{equation}
and study the large cutoff limit ($s\gg1$) of the flow equations around the Gaussian fixed point, in order to compare them with the ones obtained in section \ref{sectiongaussian}. At the leading order in$\lambda_s$, i.e. discarding the derivative of $Z_s$, the derivative of the regulator writes as:
\begin{equation}
\partial_sR_s^{\,\prime}(\vec{p})\simeq \dfrac{2}{e^{2s}Z_s}\Bigg[\dfrac{Z_s\vec{p}\,{}^2+m_s^2}{e^{(Z_s\vec{p}\,{}^2+m_s^2)/Z_se^{2s}}}\Bigg]^2\big(1-e^{-(Z_s\vec{p}\,{}^2+m_s^2)/Z_se^{2s}}\big)^2.
\end{equation}
Following the same procedure as in section \ref{sectiongfe}, we find the three equations:
\begin{align}
\partial_sm^2_s\simeq-\frac{4\lambda_s}{Z_s}\sum_{\vec{p}\in \mathbb{Z}^6}e^{-2s}e^{-\frac{Z_s\vec{p}\,{}^2+m_s^2}{e^{2s}}}\delta\Big(\sum_ip_i\Big)\times \sym\mathcal{W}_{\vec{p},\vec{p},\vec{0},\vec{0}},
\end{align}
\begin{align}
\nonumber \partial_sZ_s\simeq-\frac{4\lambda_s}{Z_s}\frac{d}{dk^2}&\Bigg[\sum_{\vec{p}\in \mathbb{Z}^6}e^{-2s}e^{-\frac{Z_s\vec{p}\,{}^2+m_s^2}{e^{2s}}}\delta\Big(\sum_ip_i\Big)\times \sym\mathcal{W}_{\vec{p},\vec{p},\vec{k},\vec{k}}\Bigg]_{k=0}
\end{align}
\begin{align}
\partial_s\lambda_s\simeq\frac{8\lambda_s^2}{Z_se^{2s}}\sum_{\vec{p}\in \mathbb{Z}^6}\dfrac{e^{-\frac{Z_s\vec{p}\,{}^2+m_s^2}{Z_se^{2s}}}-e^{-\frac{Z_s\vec{p}\,{}^2+m_s^2}{Z_se^{2s}/2}}}{Z_s\vec{p}\,{}^2+m_s^2}\delta_{p_i,0}\delta\Big(\sum_ip_i\Big).
\end{align}
For each of these equations the leading order can be easily extracted. Using the following sum formula :
\begin{align}\label{goodsum}
\sum_{n\in\mathbb{Z}}&e^{-\alpha n^2}e^{i\beta n}=\left(\dfrac{\pi}{\alpha}\right)^{1/2}\sum_{n\in \mathbb{Z}}e^{-\frac{|\beta+2\pi n|^2}{4\alpha}}=\left(\dfrac{\pi}{\alpha}\right)^{1/2}e^{-\frac{\beta^2}{4\alpha}}\left(1+2\sum_{n > 0}e^{-\frac{\pi^2n^2}{\alpha}} \cosh{\left(\dfrac{\pi n \beta}{\alpha}\right)}\right),
\end{align}
the Fourier decomposition of $\delta\Big(\sum_ip_i\Big)$ and the distributional identity:
\begin{equation}\label{gooddev}
e^{-\beta^2/4\alpha}=\sum_{n}\dfrac{\sqrt{4\pi}}{n!}[\alpha^{1/2}]^{2n+1}\delta^{(2n)}(\beta). 
\end{equation}
we find:
\begin{align}
\partial_sm^2_s\simeq-\frac{24\pi^2}{\sqrt{5}}\frac{\lambda_s}{Z_s^2}e^{2s},
\end{align}
\begin{align}
\eta(s)\simeq\frac{24\pi^2}{5\sqrt{5}}\frac{\lambda_s}{Z_s^2}\label{wf},
\end{align}
\begin{align}
\partial_s\lambda_s\simeq\frac{4\pi^2}{\sqrt{5}}\frac{\lambda_s^2}{Z_s^2}\label{coupling}.
\end{align}
Using both the equations \eqref{wf} and \eqref{coupling}, we find for the effective coupling $\lambda_s/Z_s^2:={\lambda}^r_{s}$:
\begin{equation}
\partial_s{\lambda}^r_s\simeq-\frac{28\pi^2}{5\sqrt{5}}{\lambda}_s^{r\,2},
\end{equation}
in agreement with our results of section \ref{sectiongaussian}, and with the universality argument. We will see in the next appendix that a different computation of the one-loop beta function, using yet another regularization, gives the same result.

\section{One loop computation of beta function in Schwinger regularization}

In this appendix we\label{app2} give some details about the perturbative calculation of the beta function at one-loop order. \\

$\bullet$ \textbf{2-points function}. The one particle irreducible (1PI) 2-points function divergences come from the melonic diagram depicted in figure \ref{fig12}, which contains five internal faces. From the definition of the divergence degree in section \eqref{sectiondim}, we find that the divergence degree of such a diagram is equal to $2$, implying that a wave function renormalization is needed to obtain a complete renormalized theory. Including symmetry factors, its expression writes as, in momentum representation:
\begin{equation}
\Sigma_{melo}(\vec{p})=-2\lambda\sum_{i=1}^6\sum_{q_j,\,j\neq i}\dfrac{\delta\big(\sum_iq_i\big)}{\vec{q}^2+m^2}\Big|_{q_i=p_i}.
\end{equation}
In order to extract the divergent part of this sum, we use the Schwinger representation, which consist in a cutoff in the $\alpha$ integration,
\begin{equation}
\Sigma_{melo}^{\Lambda}(\vec{p}):=-2\lambda\sum_{i=1}^6\sum_{q_j,\,j\neq i}\int_{-\pi}^{\pi}\frac{d\beta}{2\pi}\int_{1/\Lambda^2}^{+\infty}d\alpha\, e^{-\alpha(\vec{q}^2+m^2)}e^{i\beta (\sum_{j}q_j)}\Big|_{p_i=q_i}.
\end{equation}
Using again \eqref{goodsum} and \eqref{gooddev}, we find, at leading order in $\Lambda$:
\begin{equation}
\Sigma_{melo}^{\Lambda}(\vec{p})=-\frac{12\pi^2\lambda}{\sqrt{5}}\Lambda^2+\dfrac{24\pi^2\lambda}{\sqrt{5}}m^2\ln(\Lambda)+\frac{24\pi^2\lambda}{5\sqrt{5}}\ln(\Lambda)\vec{p}\,{}^2+\mathcal{O}(1/\Lambda).
\end{equation}
The first two terms correspond to a mass renormalization term, and we ignore them here. The third one is more interesting for us, as it gives the wave function renormalization  $Z:=1+\delta_Z$, with:
\begin{equation}\label{waveren}
\delta_Z=\frac{24\pi^2\lambda}{5\sqrt{5}}\ln(\Lambda).
\end{equation}

\begin{center}
\includegraphics[scale=1.4]{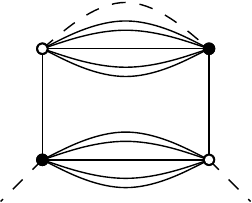} 
\captionof{figure}{Melonic contribution to the 1PI 2-points function at one loop order}\label{fig12}. 
\end{center}

$\bullet$ \textbf{4-points function}. The first one loop corrections to the 1PI 4-point function involve three types of diagrams, but only those of the type depicted in figure \ref{fig13} contribute here. As for the tadpole of the previous calculation, they involve five internal faces, but a line more than tadpole graphs. Hence, the power counting of these graphs gives a logarithmic divergence. The structure of the 1PI melonic 4-points function $\Gamma^{(4)\,melo}_{\vec{p}_1,\vec{p}_2,\vec{p}_3,\vec{p}_4}$is the following:
\begin{equation}
\Gamma^{(4)\,melo}_{\vec{p}_1,\vec{p}_2,\vec{p}_3,\vec{p}_4}=\sum_{i=1}^6\big[-4\lambda+\Delta\Gamma^{(i)}(p_{4i})\big]\frac{1}{2}\sym\mathcal{W}^{(i)}_{\vec{p}_1,\vec{p}_2,\vec{p}_3,\vec{p}_4},
\end{equation}
where:
\begin{equation}
\Delta\Gamma^{(i)}(p):=8\lambda^2\sum_{q_j,\,j\neq i}\dfrac{\delta\big(\sum_iq_i\big)}{\big(\vec{q}^2+m^2\big)^2}\Big|_{q_i=p}.
\end{equation}
Making use once again of \eqref{gooddev} and \eqref{goodsum}, we obtain 
\begin{equation}
\Delta\Gamma^{(i)\,\Lambda}(p)=\dfrac{16\pi^2\lambda}{\sqrt{5}}\ln(\Lambda)+\mathcal{O}(1/\Lambda)
\end{equation}
and, at the leading order in $\Lambda$:
\begin{equation}\label{4points}
\Gamma^{(4)\,melo}_{\vec{p}_1,\vec{p}_2,\vec{p}_3,\vec{p}_4}=-4\lambda\bigg[1-\dfrac{4\pi^2\lambda}{\sqrt{5}}\ln(\Lambda)\bigg]\sum_{i=1}^6\frac{1}{2}\sym\mathcal{W}^{(i)}_{\vec{p}_1,\vec{p}_2,\vec{p}_3,\vec{p}_4}.
\end{equation}

\begin{center}
\includegraphics[scale=1.4]{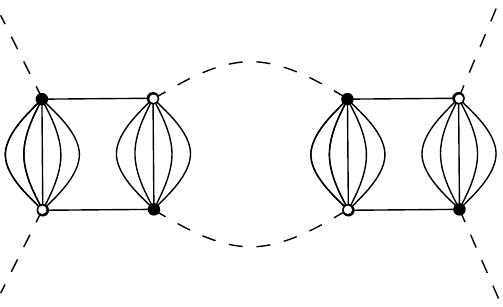} 
\captionof{figure}{Melonic contribution to the 1PI 4-points function at one loop order}\label{fig13}
\end{center}

$\bullet$\textbf{Beta function}. The previous equations \eqref{4points} and \eqref{waveren} give the renormalization of the coupling and the field at the one-loop order. The divergences are canceled by the rescaling:
\begin{equation}\label{rencouplandwave}
\lambda=Z^{-2}Z_{\lambda}^{-1}\lambda_r\qquad,\quad \psi=Z^{1/2}\psi_r
\end{equation}
with $Z$ given at order $\lambda_r$ by \eqref{waveren}, after replacing $\lambda$ with $\lambda_r$, and $Z_{\lambda}$ by:
\begin{equation}
Z_{\lambda}=1-\dfrac{4\pi^2\lambda_r}{\sqrt{5}}\ln(\Lambda)+\mathcal{O}(\lambda_r^2).
\end{equation}
Expanding the first relation \eqref{rencouplandwave}, we find, at order $\lambda_r^2$:
\begin{align}
\lambda&\simeq\lambda_r\dfrac{1+\frac{4\pi^2\lambda_r}{\sqrt{5}}\ln(\Lambda)}{1+\frac{48\pi^2\lambda_r}{5\sqrt{5}}\ln(\Lambda)}\approx \lambda_r-\dfrac{28\pi^2\lambda_r^2}{5\sqrt{5}}\ln(\Lambda)\\\nonumber
&\approx\dfrac{\lambda_r}{1+\dfrac{28\pi^2\lambda_r}{5\sqrt{5}}\ln(\Lambda)}
\end{align}
implying:
\begin{equation}
\Lambda\dfrac{d\lambda}{d\Lambda}=-\dfrac{28\pi^2\lambda_r^2}{5\sqrt{5}}=:\beta(\lambda).
\end{equation}
Again this result is consistent with those obtained in section \ref{sectiongaussian} and in appendix \ref{app1}. 
\pagebreak



\end{document}